%% file: main.tex
\documentclass[table]{biophys-mod}
\usepackage[utf8]{inputenc}
\usepackage{graphicx}
\usepackage[version=3]{mhchem}
\usepackage{array}
\usepackage{amsfonts}
\usepackage{booktabs}
\usepackage{rotating}
\usepackage{outlines}
\usepackage{multirow}
\usepackage{mhchem}
\usepackage{verbatim}
\usepackage{wrapfig}
\usepackage[colorlinks,allcolors=cyan!70!black]{hyperref}
\usepackage{natbib}
\usepackage{ccaption}

\definecolor{light-gray}{gray}{0.9}

\title{Redox Poise during \textit{Rhodospirillum rubrum} Phototrophic Growth Drives Large-scale Changes in Macromolecular Synthesis Pathways}
\runningtitle{Redox and Large-scale Changes in Macromolecular Synthesis}

\author[1,2*]{William R. Cannon}
\author[3]{Ethan King}
\author[4]{Katherine A. Huening}
\author[4]{Justin A. North}
\runningauthor{Cannon, King, Huening and North} 

\affil[1]{Computational Mathematics Group, Pacific Northwest National Laboratory, Richland, WA}
\affil[2]{Interdisciplinary Center for Quantitative Modeling in Biology, University of California, Riverside, CA}
\affil[3]{Artificial Intelligence \& Mathematical Modeling Group, Pacific Northwest National Laboratory, Richland, WA}
\affil[4]{Department of Microbiology, The Ohio State University, Columbus, OH}

\corrauthor[*]{william.cannon@pnnl.gov}

\papertype{Article}

\begin{document}
\begin{frontmatter}

\begin{abstract}
During photoheterotrophic growth on organic substrates, purple nonsulfur photosynthetic bacteria like \textit{Rhodospirillum rubrum} can acquire electrons by multiple means, including oxidation of organic substrates, oxidation of inorganic electron donors (e.g. H$_2$), and by reverse electron flow from the photosynthetic electron transport chain. These electrons are stored in the form of reduced electron-carrying cofactors (e.g. NAD(P)H and ferredoxin). The ratio of oxidized to reduced redox cofactors (e.g. ratio of NAD(P)+:NAD(P)H), or 'redox poise` is difficult to understand or predict, as are the  cellular processes for dissipating these reducing equivalents. Using physics-based models that capture mass action kinetics consistent with the thermodynamics of reactions and pathways, a range of redox conditions for heterophototrophic growth are evaluated, from conditions in which the NADP+/NADPH levels approach thermodynamic equilibrium to conditions in which the NADP+/NADPH ratio is far above the typical physiological values. Modeling results together with experimental measurements of macromolecule levels (DNA, RNA, proteins and fatty acids) indicate that the redox poise of the cell results in large-scale changes in the activity of biosynthetic pathways. Phototrophic growth is less coupled than expected
to producing reductant, NAD(P)H, by reverse electron flow from the quinone pool. Instead, it primarily functions for  ATP
production (photophosphorylation), which drives reduction even when NADPH levels are relatively low compared to NADP+.
The model, in agreement with experimental measurements of macromolecule ratios of cells growing on different carbon substrates, indicates that the dynamics of nucleotide versus lipid and protein production is likely a significant mechanism of
balancing oxidation and reduction in the cell.
\end{abstract}

\end{frontmatter}

\section*{Introduction}
Anoxygenic Purple Nonsulfur Photosynthetic Bacteria (PNSB) couple the production of high potential electrons in the light harvesting complex with the synthesis of ATP through generation of a proton motive force via bacteriochlorophyll bc$_1$ complex and the Quinone (Q) cycle (cyclic photophosphorylation) \cite{Klamt2008}. Therefore, unlike in plants and algae that perform non-cyclic photophosphorylation, PNSB do not generate reducing equivalents through water-splitting to form NADPH. Rather in PNSB, NAD(P)H reducing equivalents are ultimately acquired from the oxidation of inorganic compounds, primarily hydrogen H$_2$ and to a lesser extent sulfide (HS$^-$), thiosulfate (S$_2$O$_3^{2-}$), and carbon monoxide (CO) \cite{hansen1972, maness_2001}, and the oxidation of organic substrates. In addition, reducing equivalents that make their way to the Q pool for cyclic photophosphorylation can be recovered to regenerate NAD(P)H through reverse electron flow \cite{Klamt2008}. The reducing equivalents, in addition to cyclic photophosphorylation, are also used for anaerobic respiration or reduction of relatively oxidized carbon-containing compounds to form biomass. Most notably in PNSB, carbon dioxide (CO$_2$) can be captured and reduced by the Calvin-Benson-Bassham (CBB) cycle, whereby one turn of the cycle assimilates one CO$_2$ in an ATP-dependent manner,
\begin{align}
\text{D-ribulose-1,5-bisphosphate + CO$_2$ + H$_2$O} & \rightleftharpoons ... \rightleftharpoons \text{2 1,3-bisphospho-D-glycerate} 
\end{align}
and subsequently reduces the oxidized carbon via a hydrogenation reaction,
\begin{align}
 \text{2 1,3-bisphospho-D-glycerate + 2 NADPH} & \rightleftharpoons   \text{2 D-glyceraldehyde-3-phosphate + 2 NADP+.}
\end{align}
For example, disrupting the activity of the CBB cycle in the PNSB \textit{Rhodobacter sphaeroides} (\textit{R. sph}) prevents photoheterotrophic growth on organic substrates that are more reduced than the oxidation state of biomass, and also prevents growth on many substrates more oxidized than biomass (Table \ref{tab:oxidation_states}) \cite{falcone_1991, hallenbeck_1990}. Mechanistically, this is because assimilation of most organic substrates follows oxidative pathways that initially generate NAD(P)H. In CBB deletion strains, disposal of the reducing equivalents via reduction of CO$_2$ by the CBB cycle is no longer possible. As a result, the NAD(P)+ pool can become depleted by these oxidative processes, preventing them from proceeding forward \cite{McKinlay_2010}. Thus, \textit{R. sph} RubisCO deletion strains are unable to grow unless alternative reductive CO$_2$ fixation pathways are sufficiently available (e.g. the Arnon-Buchanan reverse TCA cycle), adequate alternative electron acceptors are provided (e.g. DMSO and TMAO), or H$_2$ can be produced via nitrogenase   \cite{gordon_JBacter_2014, wang_JBacter_2011}.

For photoheterotrophic metabolism, the main carbon sources are short chain fatty acids and other small organic acids such as acetate, propionate, butyrate, fumarate, and malate. When the NAD(P)+ pool becomes  depleted [i.e. excess NAD(P)H accumulates], organic substrates which are more oxidized than cell biomass could potentially serve as electron acceptors (Table \ref{tab:oxidation_states}) (e.g. fumarate for fumarate reductase and oxaloacetate for malate dehydrogenase). It would be an obvious solution to simply take in more of the organic acids and reduce them toward the oxidation state of biomass to regenerate the NAD(P)+ pool. However, the uptake rate or availability of the relatively oxidized organic substrates could be limiting as an electron acceptor. Also, as shown in Table \ref{tab:oxidation_states} these carbon compounds are less oxidized than CO$_2$, and so cannot accept as many reducing equivalents as CO$_2$. When CO$_2$ is used as the electron acceptor, it can originate from the environment or be generated via cellular oxidation of organic substrates. Any CO$_2$ produced from the oxidation of organic substrates initially results in a decrease in NAD(P)+:NAD(P)H ratio. So in this case, the CO$_2$ needs to be reduced to an oxidation state that is more reduced than that of the starting organic substrate to result in a net increase in NAD(P)+:NAD(P)H ratios. Given this metabolic versatility in generating and using reducing equivalents, severals questions arise: 1) what is the overall redox poise of the cell [i.e. ratio NAD(P)+:NAD(P)H and other cellular redox carriers], 2) by what reductive processes do PNSB use to dispose of reducing equivalents when they are sufficiently available or even in excess, and 3) how does this change the production of cellular macromolecular components like proteins, nucleotides, and lipids?

Understanding how \textit{R. rubrum} maintains electron balance, or redox poise, and builds cell mass during photoheterotrophic growth \cite{McCulley2020} is consequently important for future understanding on employing and engineering \textit{R. rubrum} and other PNSB for biotechnology purposes, such as hydrogen, biofuels, polyhydroxyalkonate bioplastic and ethylene bioplastic production. This is because these products have oxidation states as reduced or more reduced that cell biomass, potentially competing with the reduced electron-carrier cofactor pool   \cite{north_2020, wang_2010}. The issue of redox balance and biomass production is fundamentally a question of mass balance and thermodynamics. 


\begin{table}[hbt]
    \centering
    \begin{tabular}{l|c|c}
\toprule
Compound & Formula & Carbon Oxidation State \\
\midrule
Carbon dioxide & CO$_2$ & 4 \\
Fumarate & C$_4$H$_4$O$_4$ & 1 \\
Malate &  C$_4$H$_6$O$_5$ & 1 \\
Acetate & C$_2$H$_4$O$_2$ & 0 \\
Avg Biomass & C$_4$H$_{7.16}$O$_{2.00}$N$_{0.80}$ & -0.19 \\
Ethanol & C$_2$H$_6$O & -2 \\
\bottomrule
\end{tabular}
\caption{Carbon oxidation states of compounds used for phototropic growth compared to the average carbon oxidation state and average biomass formula calculated from across multiple microbes where the carbon oxidation state Z$_C$ = (-n$_H$ + 2*n$_O$ + 3*n$_N$)/n$_C$ \cite{Roels_1980}.}
\label{tab:oxidation_states}
\end{table}

\begin{figure}[hbt]
    \centering
    \includegraphics[width=0.45\linewidth]{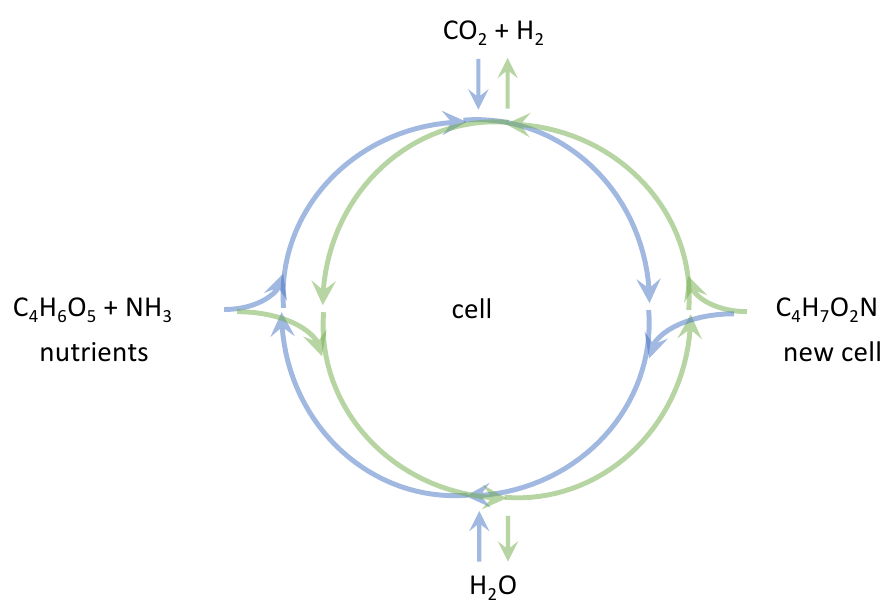}
    \caption{An abstraction of the cell cycle looking at the overall chemical reaction in which the nutrients C$_4$H$_6$O$_5$ and NH$_3$ are turned into biomass, C$_4$H$_7$O$_2$N. The cycle can operate in either the reductive (blue) or the oxidative (green) direction. Metabolism produces water as a by-product of hydrocarbon oxidation in many systems. In contrast, in reductive systems such as oxygenic non-cyclic photosynthesis, cells may consume water.}
    \label{fig:dissipative_cell}
\end{figure}

Redox balance in \textit{R. rubrum} can be understood thermodynamically in terms of the abstract cell cycle depicted in Figure \ref{fig:dissipative_cell}. In this cycle, nutrients -- malate and ammonia in this case -- are taken in and biomass with a particular elemental composition (e.g. C$_4$H$_7$O$_2$N) is produced. Even with a compound such as malate that is relatively reduced compared to CO$_2$, whether the cycle operates clockwise resulting in a net assimilation of CO$_2$ or operates counterclockwise resulting in a net production of  CO$_2$ depends on the amount of the relatively reduced compounds that are available to generate reducing equivalents (e.g. H$_2$ in this idealized case of Figure \ref{fig:dissipative_cell}). For instance, if enough NAD(P)H reductant can be made, then the CBB cycle or the reductive TCA cycle (Arnon-Buchanan cycle) can operate, assimilating CO$_2$ and producing the reduced compound, 3-phosphoglycerate for cell biomass or bioproduct synthesis \cite{wang1993}. Alternately, if reductant is not available, the idealized cell cycle of Figure \ref{fig:dissipative_cell} would operate in the counterclockwise direction, producing CO$_2$ and reductant to build biomass. In physics, such cyclical processes are known as dissipative structures.

Dissipative structures form when the need to dissipate energy forces the material to adapt a dynamic pattern in which material movement becomes correlated, typically forming cycles, that dissipate the energy the fastest way possible. Hurricanes and tornadoes are well known examples of dissipative structures. In tornado formation, the difference in heat between the earth's hot surface and the cool atmosphere causes air movement (wind) to dissipate the excess energy, in accordance with the second law of thermodynamics. If the energy difference is great enough between the ground and the upper atmosphere, the winds become correlated and cyclical. The cyclical structure of the tornado maximizes the dissipation of energy. In thermodynamics, this is known as maximizing the entropy production. (Physical constraints may prevent the system from reaching a global maximum, however.) The cyclical structure maintains mass balance in the system, otherwise all the warm air would simply move upward and obtain heat balance but not mass balance. Consequently, the formation of the highly organized, cyclical structure of the winds is a direct consequence of maximizing the entropy production rate. While entropy is often characterized as the state of disorder of a system, this is misleading and technically incorrect. Entropy, whether thermodynamic or information entropy, is simply related to the logarithm of the probability of the system to exist in a particular state \cite{Planck1914}. The highly organized cyclical structure of hurricanes and tornadoes is the most probable structure of the winds given the highly non-equilibrium conditions.

Phototrophic biological cells act as dissipative structures. Energy coming in as sunlight is captured as high potential electrons that can then be used to generate ATP via generation of a protonmotive force. For many phototrophs, this energy is dissipated in part by reducing CO$_2$ to biomass in an ATP-requiring manner. During this process, the CBB cycle acts as a dissipative cycle. Dissipative cycles occur throughout the cell, from the CBB and Arnon-Buchanan cycles of CO$_2$ assimilation \cite{Bar-Even_2012}, to the oxidative TCA cycle and many others (\textit{e.g.,} see Bar-Even \textit{et al.} \cite{Bar-Even_2012}). Whether flux through the TCA cycle is thermodynamically favored in the oxidative (clockwise) direction to oxidize metabolites to CO$_2$ and produce NAD(P)H or in the reductive (counterclockwise, Arnon-Buchanan cycle) direction to assimilate CO$_2$ and utilize NAD(P)H depends on the concentration of redox carriers in the system (i.e. NAD(P)+:NAD(P)H ratio).

During growth on malate, isotope labeling studies have been used with models to infer the direction of flux in the central metabolism of \textit{R. rubrum} \cite{McCulley2020}. The models indicated that the isotope pattern was most similar to a model in which malate entered into carbon metabolism via the TCA cycle as expected, but then flowed through reactions in opposite redox directions, both oxidatively to oxaloacetate and reductively to succinate.  
If sufficient reducing equivalents were present, then one would expect malate to proceed primarily reductively through the Arnon-Buchanan cycle. Thus, understanding and predicting the dynamics of redox carrier concentration (i.e. redox balance) and carbon flow is a challenge for PNSB, and has remained largely an open question. 


A similar issue of how redox balance is maintained and which direction of carbon flows exists for photoheterotrophic growth of \textit{R. rubrum} on acetate and ethanol. Acetate and ethanol are assimilated as acetyl-CoA and then the ethylmalonyl-CoA pathway condenses 2 molecules of acetyl-CoA followed by reduction to malate and succinate, which enter the TCA cycle and are used to form biomass.  Again, provided that sufficient reducing equivalents are available, one might intuitively expect that a significant percentage of the malate and succinate would proceed through the TCA cycle in the reductive direction (Arnon-Buchanan cycle).


In this report, we take advantage of new developments regarding the use of thermodynamics and optimal control concepts to model metabolism \cite{Britton2020, King_frontiersSysBio_2023} to predict in \textit{R. rubrum} how redox carrier ratios and biomass composition are affected when 1) the carbon sources for growth, malate and acetate, are less oxidized than carbon dioxide, and 2) the environment the cell is living in is relatively oxidizing or reducing depending on the available concentration of electron donors (e.g. H$_2$ and in Figure \ref{fig:dissipative_cell})  . A range of redox conditions are evaluated, from conditions in which the NADP+/NADPH levels are such that the pair are not far above thermodynamic equilibrium, to redox levels found physiologically, to redox levels in which NADP+/NADPH is far above physiological levels. All other redox pairs are coupled to NADP+/NADPH  through metabolism. 
The modeling results and experimental evaluation of the mass ratios of DNA, RNA, protein and fatty acids indicate that the growth substrate and resulting redox poise of the cell drives large changes in biosynthetic pathways. Furthermore, the simulation results, combined with experimental data from the literature, imply that anaerobic photoheterotrophic growth in \textit{R. rubrum} is actually not very reductive, but lies more on the oxidative end of the range.
Growth is less coupled than expected to producing reductant (NAD(P)H) by reverse electron flow from the quinone pool. Instead, growth is mostly due to high ATP production (photophosphorylation), which drives reduction even when NAD(P)H levels are relatively low compared to NAD(P)+. Furthermore, the simulations imply that the dynamics of nucleotide versus protein production may be a significant mechanism of balancing oxidation and reduction in the cell.

\section*{Methods}

\subsection*{Maximum Entropy Production Differential Equations}
Here the method for formulating the mass action differential equations that describe the equations of motion of the system of coupled chemical reactions is described.
Consider a reaction involving $n_A$ molecules of reactants $A$, and $n_B$ molecules of products $B$, each with respective unsigned stoichiometric coefficients $\nu_{i,j}$ for each molecular species, and mass action rate parameters $k_1$ and $k_{-1}$,
\begin{equation}\label{reaction_basic}
\ce{\nu_{A,1}n_{A}  <=>[k_{1}][k_{-1}] \nu_{B,1}n_{B} }.
\end{equation}
The mass action rate law is,
\begin{equation}\label{kinetic_law}
J_{net} = k_{1}n_{A}^{\nu_{A,1}}  - k_{-1}n_{B}^{\nu_{B,1}}.
\end{equation}
The mass action rate parameters $k_j$ for each of the $Z$ reactions in the system are derived from the maximum entropy production solution to the steady state. To obtain the maximum entropy production solution, the usual mass action rate law is algebraically manipulated to explicitly include thermodynamic terms, specifically the equilibrium constants $K_1$ and $K_{-1}$ and the reaction quotients $Q_1$ and $Q_{-1}$, 

\begin{eqnarray}
    J_{net} 
    &=  &k_{-1}n_{B}^{\nu_{B,1}} 
    \left( 
    \frac{k_{1}n_{A}^{\nu_{A,1}}}{k_{-1}n_{B}^{\nu_{B,1}} }
    \right)
    - k_{1}n_{A}^{\nu_{A,1}}  
    \left(
    \frac{k_{-1}n_{B}^{\nu_{B,1}} }{k_{1}n_{A}^{\nu_{A,1}} }
    \right) \label{kinetic_law_expanded1} \\
    &= & k_{-1}n_{B}^{\nu_{B,1}} (K_1 Q_{1}^{-1}) - k_{1}n_{A}^{\nu_{A,1}} (K_{-1} Q_{-1}^{-1}) \label{kinetic_law_expanded2}.
\end{eqnarray}
Equation \eqref{kinetic_law_expanded2} is an exact representation of the law of mass action, but is now separated into thermodynamic functions ($K_1 Q_{1}^{-1}, K_{-1} Q_{-1}^{-1}$) and kinetic coefficients ($k_{-1}n_{B}^{\nu_{B,1}}, k_{1}n_{A}^{\nu_{A,1}}$). Assuming that each kinetic coefficient has the same value is equivalent to assuming that each reaction occurs on the same time scale. Although this assumption is incorrect, the assumed formulation has the advantage that it makes the energy surface convex, since the surface of the log of an exponential distribution is convex for each fixed number of particles, $N_{tot}$. The result of this approximation is the Marcelin equation \cite{Marcelin1910}, which can be used to easily find the steady state of maximum entropy production \cite{Cannon2018a, Britton2020}. The time dependence can then be added back in post-hoc such that the correct formulation of the law of mass action is once again obtained. In the maximum entropy production approximation, each reaction in the system of $Z$ sequential reactions will have equal rates that are proportional to the thermodynamic driving forces such that,  
\begin{equation}\label{flux}
    J_{1,net} = J_{2,net} =...=J_{Z,net},
\end{equation}
and, 
\begin{equation}\label{fluxes}
\begin{split}
    J_{1,net} &= K_{1}Q_{1}^{-1} - K_{-1}Q_{-1}^{-1},\\
    J_{2,net} &= K_{2}Q_{2}^{-1} - K_{-2}Q_{-2}^{-1},\\
    &... ,\\
    J_{Z,net} &= K_{Z}Q_{Z}^{-1} - K_{-Z}Q_{-Z}^{-1}.
\end{split}
\end{equation}
The set of equations above produce reaction rates such that the system has the most probable steady state system and has the maximum change in the probability of the reaction system with time \cite{cannon2023a}. 
As mentioned above, the kinetics can then be adjusted such that the specific, experimentally observed reaction rates are produced if measurements of steady state concentrations of each of the reactants and products can be obtained. Adjustment of rate constants mostly impacts the specific concentrations of the intermediate reactants and products. However, no adjustment of the reaction rate parameters can change the overall thermodynamic feasibility of the pathway if the initial reactant and final product concentrations of the pathway are fixed. 

However, systems of coupled chemical reactions such as in metabolism can easily produce chemical concentrations approaching the solubility limit \cite{Atkinson1969, Atkinson1977}, and as such control of the reactions is required to prevent the solvent from becoming viscous, preventing diffusion-reaction processes \cite{Britton2020, Heimlicher2019,Parry2014}. Therefore, metabolite concentration control is implemented as follows.

\subsection*{Control and Optimization}

Control of metabolite concentrations is implemented using activity coefficients for the reactions that reflect the activity of the enzyme catalyst. The activity coefficient for any reaction $j$ ranges from $\alpha_j = 0$ where the enzyme has no activity to $\alpha_j = 1$ where the enzyme is considered fully active. The activity coefficient exerts control over a reaction by scaling the reaction flux,
\begin{equation}\label{control1}
\begin{split}
    J_{j,net}(\alpha_j) &= \alpha_j(K_{j}Q_{1}^{-1} - K_{-j}Q_{-1}^{-1})
\end{split}
\end{equation}

While controlling metabolite concentrations may be a primary role of metabolic regulation \cite{Atkinson1969, Atkinson1977}, natural selection also requires that organisms grow fast and efficiently - using the available energy from the environment to ensure survival and compete with others. Often, this means down-regulating reactions that do not significantly increase the fitness or ability of the organism to replicate.
We use a method referred to as pathway-controlled optimization (PCO) to obtain this biological goal. We will briefly describe the intent of the method next. For full details see \cite{King_frontiersSysBio_2023}.

Let $\mathcal{G} \subset \mathcal{J}$ be the set of reactions corresponding to production of biomass, such as the reactions producing DNA, RNA, proteins and fatty acids. We formulate the steady state with maximum biomass production as the solution of the optimization problem,

\begin{subequations}
\begin{align}
& \max  \sum_{j \in \mathcal{G} } J_{j}(n, \alpha_{j} ) \\
&\mbox{subject to:} \nonumber\\
&\frac{d n_{i} }{dt} ~ =  ~0 ~ ~ \forall~ i \in \mathcal{I}_{v} \, , \label{Gr_std1} \\
&0 \leq n_{i} \leq n_{\max} ~ ~ \forall ~ i \in \mathcal{I}_{v} \, , \label{n_feas} \\
&n_{i} ~ = ~  \bar{n}_{i} ~  ~\forall ~ i \in \mathcal{I}_{f}  \, , \label{Gr_std2} \\
&0 \leq \alpha_{j} \leq 1 ~ ~ \forall ~ j \in \mathcal{J} \, . \label{a_feas}
\end{align}
\label{Gr_opt}
\end{subequations}
The objective seeks to maximize the flux through the growth reactions $\mathcal{G}$ while the constraint \eqref{Gr_std1} ensures that the steady state condition is satisfied and constraint \eqref{n_feas} ensures that the metabolites stay within their physiological or experimentally measured values. The boundary condition for the ordinary differential equations consist of fixed boundary species such as ATP, NAD(P)H, NAD(P)+, and other cofactors. The boundary conditions are enforced by constraint \eqref{Gr_std2}. Likewise, the activity coefficients are constrained to values $[0,1]$ by constraint \eqref{a_feas}.

The formulation of the problem is simple to express but difficult to solve. The steady state constraints \eqref{Gr_std1} are nonlinear and non-convex presenting significant challenges to optimization. Values for the flux, activity coefficients, and metabolite concentrations can also vary over many orders of magnitude, which introduces additional difficulty in employing numerical methods to compute solutions. We reformulate the optimization problem to be more computationally tractable and solve the problem with an interior point solver \cite{WachterIpopt} and advanced linear algebra library \cite{HSL_library}, as descibed in \cite{King_frontiersSysBio_2023}. Open source code in the form of a Jupyter Notebook in python is freely available (Supplementary Material.)

As the system approaches the optimum, the optimization surface can become flat and the optimization can have difficulties converging in some cases. This specifically can be an issue as some reactions in the system approach equilibrium while other reactions are far from equilibrium. In these cases, some reaction free energies and their respective fluxes both approach 0.0, making control of the reactions, and hence convergence, effectively noisy. 

In our experience, convergence can always be obtained by adjusting either the hyperparameter controlling lower bounds on metabolite log concentrations, $VarM\_lbound$ in the source code, or the hyperparameter $Mb$, which scales the reaction free energies, as these hyperparameters manipulate the optimization surface. However, in order to be able to compare steady state solutions across different redox conditions, we always used $VarM\_lbound = -300$ and $Mb = 1000$, and in cases in which the optimizations did not converge, we adjusted the redox condition minimally by increasing or decreasing the ratio of NADP+/NADPH to be slightly above or below the corresponding target level for $KQ^{-1}$ (shown in the top row of Table \ref{tab:nad_nadph_odds}). 
Specifically, see the top of the rightmost column in Table \ref{tab:nad_nadph_odds}, in which case the target for the reaction odds $KQ^{-1} = 10^{-3}$ but instead of using the corresponding value for the NADP+/NADPH ratio of $7.38 \time 10^{-4}$, a value of $7.57 \times 10^{-4}$ was used, which corresponds to an actual $KQ^{-1} = 0.975\times 10^{-3}.$ The optimizations were carried out on an Apple MacBook Pro with an Intel i9 core and 64 GB of memory.

\subsection*{Metabolic Model}
The model consists of 252 reactions and 253 metabolites. Reaction pathways include those important for \textit{R. rubrum} photoheterotrophic growth, including all of the reactions of central metabolism, the ethylmalonyl-CoA pathway for ethanol/acetate assimilation. Secondary metabolism included reactions for amino acid and nucleic acid synthesis, folate metabolism, S-adenosyl methionine metabolism, protein synthesis, DNA synthesis, RNA synthesis and degradation. The growth reaction for biomass included protein synthesis, DNA, RNA, and fatty acid synthesis according to the relative abundances measured experimentally under fumarate growth conditions \cite{McCulley2020}. The metabolic model was developed using Pathway Tools \cite{Karp2016} and stored as a BioCyc pathway genome database \cite{Karp2019}. The model is available as a Jupyter notebook in python as described in the Supplemental Material.
    
Of the total 253 metabolites, 227 of the metabolites are free variables and 26 metabolites are fixed as boundary conditions. The rank of the stoichiometric matrix is 236. The set of fixed metabolites (boundary conditions) and their concentrations are listed in Table \ref{tab:fixed_metabolites} (Supplementary Material). In addition, standard free energy of reaction for the ATP synthesis reaction was adjusted to account for a 10-fold driving force due to the proton gradient across the cytoplasmic membrane.

Each reaction is modeled using mass action kinetics, as described above. Equilibrium constants for the reactions were determined using the eQuilibrator API software, version 0.4.0 \cite{Noor2013}, except where noted in the Supplementary Material. A pH of 7.0 and ionic strength of 0.15 M were used throughout. 

The steady state solutions were constrained to replicate the experimentally observed ratios of DNA:RNA:protein:lipid during growth on malate obtained by McCulley et. al of 1.0: 2.9: 44.1 previously \cite{McCulley2020}, which agree with those measured in this work for malate (Table \ref{tab:macro_ratios}). This was done by incorporating these amounts of the respective macromolecules into biomass, in the form of a cell monomer, using the following pseudo-chemical reaction equation,
\begin{align}
    \text{1.0 DNA + 2.9 RNA + 44.1 protein + 8.5 lipid} \rightleftharpoons \text{1.0 cell monomer}
    \label{macromolecule_constraint}
\end{align}
While the ratio of protein:RNA:DNA will differ experimentally as a function of the nutrient, the use of a fixed value calibrated to malate growth allows for the comparison of the energetics and dynamics of growth for the different nutrients for the same growth process. For fatty acid composition, we assumed that the fatty acid mass percentage of a \textit{R. rubrum} cell is 15\% and that palmitate could be used to represent fatty acid (FA) in the model. Palmitate has a molecular mass of 256.42 Da. 
Accordingly, the total macromolecule mass estimate relative to DNA is 
DNA + RNA+ Protein + FA mass = 1.0 + 2.9 + 44.1 + x such that
the fatty acid mass percentage x was estimated from
0.15 = x / (1.0 + 2.9 + 44.1 + x),  
yielding x = 8.47\%. Subsequently, for comparison of modeling to experimental results, the macromolecule content and elemental composition of \textit{R. rubrum} growing on various substrates were measured as detailed in the experimental methods section. 

\hfill \break
\textit{Predicted Biomass elemental composition.} The predicted elemental composition of the modeled biomass formation is then found by the following procedure. The overall chemical equation for the cell is determined from the rate of production and consumption of the fixed (boundary) metabolites, normalized to a rate of malate or acetate consumption of 1000 molecules. However, if a normalized rate is less than 1/100th of the rate of malate, then it is not included in the overall chemical reactions (Tables \ref{tab:overall_malate} and \ref{tab:overall_acetate}). Using the overall reaction from a simulation (middle row of Table \ref{tab:overall_malate}) as an example, the chemical equation is,
\begin{align*}
    \text{2.2 THF + 1.0 \ 5,10-MTHF + 0.06 5-MTHF + 0.08 SO$_4$ + 1.85 ABP + 18.5 P$_i$} 
    \text{+ 10 C$_4$H$_6$O$_5$ + 5.32 NH$_3$} \\
    \text{+ 0.7 NAD+ + 30.7 NADPH + 0.1 CO$_2$} \rightarrow \text{3.3 N10-fTHF + 4.06 P$_{i,2}$ +  biomass + 31.4 NADP+}
\end{align*}
First, the internal metabolites except phosphate are dropped such that the overall equation is, 
\begin{align}
\text{0.7 NAD+ + 19.2 P$_i$ + 10 C$_4$H$_6$O$_5$ + 5.32 NH$_3$ + 30.70 NADPH + 0.1 CO$_2$} & \rightarrow \text{biomass + 31.4 NADP+.}
\end{align}
Next, the chemical equation is balanced for redundant redox components NAD+ and NADP by adding,
\begin{align}
    \text{0.7 NADP} & \rightarrow \text{0.7 NAD+ + 0.7 P$_i$}
\end{align}
and canceling like terms so that the overall chemical equation becomes,
\begin{align}
  \text{18.5 P$_i$ + 10 C$_4$H$_6$O$_5$ + 5.32 NH$_3$ + 30.7 NADPH + 0.1 CO$_2$}  & \rightarrow  \text{biomass + 30.7 NADP+.}   
\end{align}
To calculate the CHON biomass elemental composition, phosphate is dropped, NADPH is replaced by H$_2$, NADP+ is replaced by a variable amount of H$_2$O. Next, 'biomass' is substituted by the variable composition formula C$_4$H$_{u}$O$_{2.04}$N$_z$, where the value of 2.04 for oxygen was determined from experimental elemental analysis of biomass (for biomass grown on acetate, a value of 1.57 was used \cite{Favier_2003}) (Table 6). The resulting chemical equation,
\begin{align*}
   \text{10 C$_4$H$_6$O$_5$ + 5.32 NH$_3$ + 30.7 H$_2$ + 0.1 CO$_2$}
   \rightarrow  \text{x C$_4$H$_{u}$O$_{2.04}$N$_z$ + y H$_2$O}, 
\end{align*}
can be solved for the values of x, y, z and u by balancing carbon, oxygen, nitrogen and hydrogen, respectively \cite{Doran2013}.

\subsection*{Experimental quantification of macromolecules and element ratios}

\textit{R. rubrum} ATCC 11170 was grown as previously reported in 500 ml of Ormerod's minimal media in sealed Roux bottles at 30 degrees Celcius under 2000 lux incandescent illumination in biological triplicate. Cultures were supplemented with 85 mM ethanol and 0.1\% mM sodium bicarbonate, 10 mM sodium-butyrate with 0.1\% sodium bicarbonate, 20 mM DL-malic acid, or 20 mM acetic acid and flushed with nitrogen to establish anaerobic conditions \cite{north_2020}. Cells were collected by centrifugation during exponential growth at an Optical Density at 660 nm of $\sim$0.80.

\textit{Biomass elemental analysis.} Pellets from 200 ml culture were  washed three times with ultrapure water to remove compounds present in the residual media, and lyophilized to dryness. Dry cell mass
(50 mg) was analyzed on a VarioEL Cube elemental CHNS/O analyzer (Elementar, New York) following manufacturer's protocols at the DOE EMSL laboratory.

\textit{Dry cell weight.} Pellets from 50 ml culture were lyophilized to dryness in pre-weighed test tubes and measured gravimetrically to determine the dry cell weight per volume and density of culture at time of cell collection for calculating macromolecules by mass ratio.

\textit{Lipid analysis.} Total extractable hydrophobic compounds (lipids, photosynthetic pigments, poly-hydroxy butyrate (PHB)) were measured by organic extraction following a modified version of the Folch method \cite{Castruita_plantCell_2011, FOLCH_JBC_1957}. 50 ml cell culture was collected by centrifugation in chloroform-treated plastic tubes. Pellets were lyophilized to dryness and extracted using 2 mL chloroform:methanol (2:1 [v/v]). After centrifugation at 4000 g for 5 minutes, the supernatant was transferred to a new glass tube, and the remaining pellet was re-extracted twice more with 2 mL chloroform:methanol as before. Extracts were pooled and washed with 0.9\% KCl at a ratio of chloroform:methanol:0.9\% KCl of 8:4:3 [v/v/v]. After addition of KCl, samples were vortexed, centrifuged, and the upper aqueous phase aspirated off. This was performed for a total of three times and afterward the lower organic phase was collected and transferred to a pre-weighed glass lyophilization vial. Solvent was removed by evaporation under a stream of nitrogen at 50 degrees Celcius and total extractable hydrophobic compounds were measured gravimetrically. To determine total lipids, the total measured PHB and photosynthetic pigment (chromatophores) were subtracted out as detailed below.

Total PHB present in cell pellet from 20 ml of cell culture was quantified using the method of Slepecky and Law \cite{law_jBacter_1961}. Similarly total photosynthetic pigment (chromatophores) was quantified using the method of van der Rest and Gingras by UV-Vis quantitation of extracted chromatophores in acetone \cite{VANDERREST_jbc_1974}.

\textit{Protein quantification.} Cells from 5 ml culture were resuspended in 700 $\mu$l EDTA-free buffer (0.1 M Tris-HCl, 2\% SDS, pH 8.0), then sonicated  for 3 minutes with 1 second on and 5 second off pulses. Samples were centrifuged for 3 minutes at 5,000 g and the supernatant was used in BCA assays (Pierce, Thermo Scientific) following manufacturer’s instructions. All BCA assays were performed in technical duplicate on the biological replicates.

\textit{DNA/RNA quantification.} Cells from 5 ml culture were immediately treated with Qiagen RNAprotect upon cell collection by centrifugation and stored at 4 degrees Celcius until analyzed. RNAprotect was removed by centrifugation and cells were resuspended in 700 $\mu$l of 0.1 M Tris-HCl, 2\% SDS, 0.1 M EDTA, pH 8.0. Cells were sonicated as above and centrifuged for 3 minutes at 5,000 g. HPLC analysis was performed by the method of Dell'anno \textit{et. al} \cite{dellanno_appEnvironMicro_1998} with the following modifications. Supernatants were applied to an anion exchange column (TSKgel DEAE-5PW, TOSOH Bioscience) connected to a Shimadzu Prominence HPLC with UV detection at 260 nm. DNA and RNA were eluted from the column on a gradient of 0.1-1 M KCl in 20 mM K-phosphate buffer with 5 M urea, pH 6.8, over 24 minutes. DNA and RNA concentrations were calculated from peak areas compared to standard calibration curves generated using DNA from salmon testes (Sigma-Aldrich) and RNA from baker’s yeast (Sigma-Aldrich).

\section*{Results}





As discussed above, we are interested in the redox conditions governing the assimilation of malate and acetate relative to CO$_2$ production or consumption, and how the metabolism of \textit{R. rubrum} adjusts for differing internal redox conditions that arise due to varying concentration of additional electron donors (e.g. H$_2$).  As shown in Figure \ref{fig:dissipative_cell} the overall growth reaction can operate either in the clockwise (reductive, blue) or counterclockwise (oxidative, green) direction depending on the amount of electron donors (e.g. H$_2$) and thus reducing power available to the cell. As an example, Table \ref{tab:oxidation_states_malate} lists several of these scenarios that can occur during growth on malate, depending on whether the cell environment is strongly reductive (Table \ref{tab:oxidation_states_malate}, bottom row), mildly reductive, neutral or oxidative (Table \ref{tab:oxidation_states_malate}, top row). In this context, oxidative refers to a net release of CO$_2$, and reductive refers to a net consumption.  Each scenario utilizes a different amount of reductant in the form of H$_2$, and negative values indicate consumption while positive values indicate production in Table \ref{tab:oxidation_states_malate}. The stoichiometric ratio of malate to CO$_2$ used during growth under each redox condition is shown in the third column. 
Likewise, Table \ref{tab:oxidation_states_acetate} lists similar model oxidation-reduction scenarios for growth on acetate. However, this is an oversimplified model, as it assumes the cell biomass composition (CHNO ratios) are a constant for the cell. Rather, the ability of cells to  differentially produce macromolecules (nucleotides, proteins, lipids, carbohydrates) based on the oxidation state of the specific growth substrate and reducing power available to the cell must be considered, as this will result in a different biomass composition and different overall oxidation state of cell biomass.

\bgroup
\def\arraystretch{1.5}
\begin{table}[hbt]
    \centering
    \small
    \begin{tabular}{l|rcl|c}
\toprule
Redox Condition & \multicolumn{3}{c}{Overall Reaction} & malate:CO$_2$ \\
\midrule
Oxidative & 3\ ${C_4}{H_6}{O_5} + 0.63\ {N}{H_3}$ & $\rightarrow$ & $2\ {C_4} {H_{7.07}}{O_{2.04}}{N_{0.63}} + 4\ {C} {O_2} + 3\ {H_2}{O} + 1.63\ {H_2}$ & -3:4 \\ 
Neutral (approx.) & ${C_4}{H_6}{O_5} + 0.63\ {N}{H_3} + 2.63\ {H_2}$ & $\rightarrow$ & \ \ \ ${C_4} {H_{7.07}} {O_{2.04}}{N_{0.63}} + 3.04\ {H_2}{O}  $ & -1:0 \\
Mild Reducing &  $C_4{H_6}{O_5} + 2\ {C}{O_2} + 0.95 {N}{H_3} + 6.83 \ {H_2} $ & $\rightarrow$ &  $1.5\ {C_4}{H_{7.07}}{O_{2.04}}{N_{0.63}} + 5.94\ {H_2}{O}$ & -1:-2 \\
Reducing & ${C_4}{H_6}{O_5} + 4\ {C}{O_2} + 1.26 {N}{H_3} + 11.1 \ {H_2} $ & $\rightarrow$ &  $2\ {C_4} {H_{7.07}} {O_{2.04}}{N_{0.63}} + 8.92\ {H_2}{O}$  & -1:-4 \\
\bottomrule
\end{tabular}
\caption{Model redox scenarios for growth on malate with H$_2$ and CO$_2$ production or consumption resulting in biomass of the assumed fixed formula. In the third column, negative values indicate consumption of malate or CO$_2$ and positive values indicate production. Coordinately, Redox Condition refers to the net oxidation or reductive environment the cell is in due to available growth substrate and electron donors (e.g. H$_2$) relative to biomass carbon oxidation state.}
\label{tab:oxidation_states_malate}
\end{table}
\egroup

\bgroup
\def\arraystretch{1.5}
\begin{table}[hbt]
\small
    \centering
    \begin{tabular}{l|rcl|c}
\toprule
Redox Condition & \multicolumn{3}{c}{Overall Reaction} & acetate:CO$_2$ \\
\midrule
Oxidative & 3\ ${C_2}{H_4}{O_2} + 0.63\ {N}{H_3}$ & $\rightarrow$ & ${C_4} {H_{7.07}}{O_{2.04}}{N_{0.63}} + 2\ {C} {O_2} + 1.96\ {H_2}{O} + 1.45\ {H_2}$ & -3:2 \\ 
Neutral (approx.) & $2\ {C_2}{H_4}{O_2} + 0.63\ {N}{H_3} + 0.55\ {H_2}$ & $\rightarrow$ & ${C_4} {H_{7.07}} {O_{2.04}}{N_{0.63}} + 1.96\ {H_2}{O}  $ & -2:0 \\
Mild Reducing & ${C_2}{H_4}{O_2} + 2\ {C}{O_2} + 0.63 {N}{H_3} + 2.55 \ {H_2} $ & $\rightarrow$ &  ${C_4} {H_{7.07}} {O_{2.04}}{N_{0.63}} + 1.96\ {H_2}{O}$  & -1:-2 \\
\bottomrule
\end{tabular}
\caption{Model redox scenarios for growth on acetate with H$_2$ and CO$_2$ production or consumption resulting in biomass of the assumed fixed formula. In the third column, negative values indicate consumption of acetate or CO$_2$ and positive values indicate production. Coordinately, Redox Condition refers to the net oxidation or reductive environment the cell is in due to available growth substrate and electron donors (e.g. H$_2$) relative to biomass carbon oxidation state.}
\label{tab:oxidation_states_acetate}
\end{table}
\egroup

To elucidate how the metabolism of \textit{R. rubrum} can produce a cellular redox poise (NADP+:NADPH ratio) and associated biomass production scenarios, we constructed a mass action thermodynamic and kinetic model of \textit{R. rubrum's} metabolism that included 242 metabolites and 239 reactions, as shown in Figure \ref{fig:pathway_overview}, including pathways for amino acid, nucleotide, RNA, DNA, and lipid synthesis, photosynthesis and the electron transport chain. 
The model of photosynthesis and the electron transport chain follows that outlined by Klamt, \textit{et al.} \cite{Klamt2008} for purple nonsulfur bacteria based on experimental studies \cite{brandt_1994}. Because of the importance to redox conditions in the cell, the processes of the electron transport chain will be discussed in detail. 

The degree to which the electron transport chain acts reductively depends on the ratio of the redox pair quinone (Q) and quinol (QH$_2$). Quinones are a chemical class that include specific chemical species such as ubiquinone, metaquinone, rhodoquinone and others that undergo oxidation-reduction between quinone (Q), semi-quinone (Q$^{.-}$) and quinol (QH$_2$) forms. During phototrophic growth the ratio of Q/QH$_2$ is largely determined by the rate of light harvesting reactions occurring in the photosynthetic apparatus, which consists of the light harvesting complex (LHC) and the reaction center (RC).
The light reaction has an overall stoichiometry in which two photons (h$\nu$) and a quinone oxidize two ferrocytochromes (cyt-c$^{2+}$) to produce ferricytochrome (cyt-c$^{3+}$) and a quinol using two protons from the cytoplasm (H$^+_c$),
\begin{align}
    \text{2 h$\nu$ + 2  cyt-c$^{2+}$ + Q + 2H$^+_c$} \rightarrow & \text{2 cyt-c$^{3+}$ + QH$_2$}. \label{eq:bc1_reaction1}
\end{align}
Ferricytochrome is recycled back to ferrocytochrome in the cytochrome bc$_1$ complex by a multi-step process involving the extraction of two additional protons from the cytoplasm (H$^+_c$) and the net depositing of four protons to the periplasm (H$^+_p$) with an overall stochiometry, 
\begin{align}
    \text{2 cyt-c$^{3+}$ + QH$_2$ + 2 H$^+_{c}$}  \rightarrow & \text{2  cyt-c$^{2+}$ + Q + 4 H$^+_{p}$} \label{eq:bc1_reaction2}
\end{align}
The overall process forms a cycle (Cycle 1) involving both the cytochromes and the quinone/quinol pair,
\begin{eqnarray*}
    \text{Q + 2  cyt-c$^{2+}$\ \ \ } & \rightleftharpoons  &\text{QH$_2$ + 2 cyt-c$^{3+}$ } \\
     \updownarrow\text{\ \ \ \ \ \ \ \ \ \ \ \ \ \ \ \ \ \ \ } & & \text{\ }\updownarrow\\ \text{\ \ Q + 2  cyt-c$^{2+}$ + 4 H$^+_{p}$}
     & \rightleftharpoons & \text{QH$_2$ + 2 cyt-c$^{3+}$ 2  + H$^+_{c}$} \\
    &  \text{Cycle 1}  \nonumber & 
\end{eqnarray*}
in which the cycle extracts four protons from the cytoplasm and deposits four protons into the periplasm, thus creating a proton motive force that can be utilized later to drive ATP synthesis via ATP synthase.

Separately, the quinone/quinol pair can also be involved in oxidation-reduction cycles of cytoplasmic NAD(P)+/NAD(P)H and succinate/fumarate (Cycle 2),
\begin{eqnarray}
    \text{QH$_2$ + NAD(P)+} & \rightleftharpoons  &\text{Q + NAD(P)H} \label{eq:reduceNAD}\\
     \updownarrow\text{\ \ \ \ \ \ \ \ \ \ \ \ \ \ \ \ \ \ \ } & & \text{\ }\updownarrow \nonumber \\
    \text{QH$_2$ + fumarate} & \rightleftharpoons & \text{Q + succinate} \\
        &  \text{Cycle 2} \nonumber &
\end{eqnarray}
The reaction of Eqn \ref{eq:reduceNAD} is generally unfavorable but can be aided by coupling to the proton motive force generated by the cytochrome bc$_1$ reaction in Cycle 1 \cite{brandt_1994}, shown in Eqn \ref{eq:bc1_reaction2}. If Cycle 1 and Cycle 2 are strongly coupled either through using the same quinone pool or through the proton motive force, the process of reverse electron flow can produce high levels of the reductant NAD(P)H due to the photosynthetic activity. 
The ratio of Q/QH$_2$ determines whether succinate is oxidized to fumarate, as in the oxidative TCA cycle, or whether fumarate is reduced to succinate, as in the reductive TCA cycle. As mentioned above, in the case of phototrophic growth, this ratio is thought to be largely driven by Cycle 1. During aerobic heterotrophic growth, an analogous cycle involving the oxidation of QH$_2$ by oxygen to Q and water takes the place of Cycle 1.

Although the Q/QH$_2$ ratio ultimately controls the downstream redox state of the cell, the light reactions involving cycling of quinols and quinones were not used to drive the system in the model. Instead, the NADP+/NADPH ratio was used to drive the system.
Consequently, to understand metabolism as a function of redox state of the cell, we evaluated the metabolic activity as a function of the NADP+/NADPH ratio. We evaluated a range of conditions from a low value at which the NADP+ concentration is set such that the redox equation,
\begin{equation}
    \text{NADPH $\rightleftharpoons$ NADP+ + H$_2$ \ \ \ \ \ \ $\Delta G^\circ = 35.0$ kJ/mol},
    \label{NADPH_NADP_redox_reaction}
\end{equation}
is at equilibrium (NADP+ = $8.86 \times 10^{-11}$ M, NADPH = $0.00012$ M) to higher values of NADP+ to NADPH ratios, corresponding to decreasing reductive power. These conditions are characterized by 
the thermodynamic odds of the reaction, defined as the usual product of the equilibrium constant and the reciprocal of the reaction quotient, $KQ^{-1}$. The thermodynamic odds for the NADP+:NADPH ratio used in setting the redox state was,
\begin{equation}
    KQ^{-1} = e^{-\Delta G^\circ/RT} \cdot\frac{[NADP+][H_2]}{[NADPH]},
    \label{eq:thermodynamic_odds}
\end{equation}
in which $[H_2]$ is 1 M. The correspondence between ratios of NADP+ and NADPH concentrations to thermodynamic odds are shown in Table \ref{tab:nad_nadph_odds}. Importantly to note, at a thermodynamic odds of $10^{-4}$ the NADP+/NADPH ratio is approximately $10^{-2}$, which  is the generally accepted value for the average NADP+/NADPH ratio and thus redox poise of the cell. However, the concentration ratios shown in Table \ref{tab:nad_nadph_odds} cannot be directly compared to experimentally measured values because the concentrations used in the simulation model are those of the unbound metabolic species, since it is the unbound concentrations that determine the overall thermodynamics of the reaction, while experimental assays typically measure whole-cell populations, including both enzyme-bound and unbound species. As the thermodynamic odds increase from $10^{-4}$ toward $10^{-7}$ the NADP+/NADPH ratio approaches unity indicating that the cell redox poise is relatively more oxidizing than typically accepted values, and as thermodynamic odds decrease from $10^{-4}$ toward 1.0 the NADP+/NADPH ratio becomes increasingly smaller indicating that the cell redox poise is relatively more reducing than typically expected values.  
\begin{table}[bt]
    \centering
    \begin{tabular}{|c|c|c|c|c|c|c|c|c|}
    \toprule
    KQ$^{-1}$  & $10^{-10}$ & $10^{-9}$ & $10^{-8}$ &  $10^{-7}$ & $10^{-6}$ & $10^{-5}$ & $10^{-4}$ & $10^{-3}$ \\
    \text{NADP+/NADPH} & 7383.5  & 738.35 & 73.8 & 7.38 & 0.738 & 7.38 $\times 10^{-2}$ & 7.38 $\times 10^{-3}$ & 7.57 $\times 10^{-4}$\\
    \hline
    \multicolumn{9}{|l|}{\underline{Malate Growth:}} \\
    \text{NAD+/NADH} & 1.44e+05  & 1.60e+05 & 3.97e+04 & 1.70e+05 & 1.60e+04 & 5.23e+03 & 2.97e+03 & 6.16e+02 \\
    \text{Q/QH$_{2}$} & 3.88e-10  & 6.50e-11 & 1.10e-11 & 4.78e-12 & 8.25e-13 & 1.44e-13 & 2.19e-14 & 3.91e-15 \\
    \text{Fe$_{ox}$/Fe$_{red}$} & 2.70e+06  & 1.06e+06 & 4.57e+05 & 1.51e+05 & 9.55e+04 & 4.05e+04 & 1.60e+04 & 7.07e+03 \\
    \text{Trdx$_{ox}$/Trdx$_{red}$} & 3.60e+02  & 4.78e+01 & 6.27e+00 & 8.05e-01 & 1.31e-01 & 2.14e-02 & 3.26e-03 & 5.74e-04 \\
    \text{cyt-C$_{ox}$/cyt-C$_{red}$} & 1.71e+01  & 7.01 & 2.89e+00 & 1.90e+00 & 7.90e-01 & 3.30e-01 & 1.29e-01 & 5.44e-02 \\
    \text{FAD/FADH2} & 5.08e+00  & 5.54e-01 & 1.92e-01 & 5.55e-03 & 3.39e-04 & 2.07e-05 & 1.02e-06 & 2.01e-08 \\
    \hline
    \multicolumn{9}{|l|}{\underline{Acetate Growth:}} \\
    \text{NAD+/NADH}                  & 2.82e+05  & 8.32e+04 & 6.90e+04 & 8.83e+03 & 6.74e+03 & 5.20e+03 & 2.70e+03 & 2.29e+03 \\
    \text{Q/QH$_{2}$}                 & 2.06e-12  & 2.87e-10 & 4.31e-03 & 5.17e-11 & 1.23e-15 & 3.39e-14 & 8.30e-15 & 2.15e-15 \\
    \text{Fe$_{ox}$/Fe$_{red}$}       & 5.03e+06  & 1.76e+06 & 4.92e+05 & 1.96e+05 & 6.73e+04 & 2.98e+04 & 1.26e+04 & 5.08e+03 \\
    \text{Trdx$_{ox}$/Trdx$_{red}$}   & 5.56e+02  & 7.34e+01 & 5.80e+00 & 1.04e+00 & 1.24e-01 & 1.50e-02 & 1.92e-03 & 2.53e-04  \\
    \text{cyt-C$_{ox}$/cyt-C$_{red}$} & 1.25e+00  & 1.47e+01 &5.71e+04 & 6.25e+00 & 3.05e-02 & 1.60e-01 & 7.92e-02 & 4.03e-02 \\
    \text{FAD/FADH2}                  & 9.54e+02  & 1.11e+04 & 1.99e+02 & 2.49e+03 & 3.90e+03 & 6.23e+03 & 1.56e+04 & 2.40e+04 \\
    \bottomrule
    \end{tabular}
    \caption{The top two rows show the correspondence between thermodynamic odds of NADPH reduction to NADP+ (Eqn \ref{eq:thermodynamic_odds}) and ratio of concentrations of NADP+ to NADPH. The second row shows the actual ratio of concentrations used. These values are held fixed in each simulation. Lower rows show the observed ratio of various redox pairs during each simulation.}
    \label{tab:nad_nadph_odds}
\end{table}

Malate and acetate were considered as organic carbon sources. Then under different thermodynamic odds of NADP+/NADPH (Table \ref{tab:nad_nadph_odds}), the redox carrier oxidation/reduction state ratios, nutrient uptake, growth and CO$_2$ assimilation activities were evaluated for the main routes of CO$_2$ assimilation: the CBB cycle, the ethylmalonyl CoA pathway, the ferredoxin-dependent 2-oxoglutarate synthase reaction, the isocitrate dehydrogenase reaction, the ferredoxin-dependent pyruvate synthase reaction, and the phosphoenolpyruvate carboxykinase reaction (Figure 2). The chemical equation for each of these main reactions for CO$_2$ assimilation is shown in Table \ref{tab:co2_assimilation_reactions}. 

\begin{table}[hbt]
    \small
    \centering
    \begin{tabular}{lrcl}
    \toprule
     Enzyme/Pathway    & \multicolumn{3}{c}{Reaction} \\
     \toprule
     RubisCO  & D-ribulose-1,5-bisphosphate + CO$_2$ + H$_2$O & $\rightleftharpoons$ & 2 3-phospho-D-glycerate  \\
     EthylMal CoA & ATP + H$_2$CO$_3$ + propanoyl-CoA & $\rightleftharpoons$ & ADP + (S)-methylmalonyl-CoA + P$_i$  \\
     2-oxoglutarate synthase & CO$_2$ + 2.0 Fedox(red) + succinyl-CoA & $\rightleftharpoons$ & CoA + 2.0 Fedox(ox) + 2-oxoglutarate \\
     isocitrate dehydrogenase &CO$_2$ + NADPH + 2-oxoglutarate & $\rightleftharpoons$ & NADP+ + D-threo-isocitrate\\
     pyruvate synthase & acetyl-CoA + CO$_2$ + 2.0 Fedox(red) & $\rightleftharpoons$ & CoA + 2.0 Fedox(ox) + pyruvate \\
     phosphoenolpyruvate carboxykinase & oxaloacetate + GTP & $\rightleftharpoons$ & CO2 + phosphoenolpyruvate + GDP \\
    \bottomrule
    \end{tabular}
    \caption{Primary CO$_2$ assimilation reactions and enzyme catalysts.}
    \label{tab:co2_assimilation_reactions}
\end{table}

The observed fluxes through these reactions, which are a direct manifestation of the thermodynamics, are shown in Figure \ref{fig:CO2vNADP}. The values on the horizontal axis (x axis) represent the thermodynamic odds: the ratio of NADP+/NADPH equilibrium constant (K) relative to the reaction quotient (Q) for NADP+/NADPH found for Eqn \ref{NADPH_NADP_redox_reaction}. That is, a value of 1.0 indicates the reaction NADP+:NADPH ratio has gone to equilibrium, whereas at a thermodynamic odds of $10^{-4}$ the NADP+/NADPH ratio is held at a value 1000-fold greater than the equilibrium ratio.  

For both malate and acetate, several similar trends stand out. First, as the redox level moves away from typical values (thermodynamic odds of $10^{-4}$) towards more oxidative conditions, the overall level of 
CO$_2$ assimilation decreases, as expected. Second, CO$_2$ assimilation through the CBB cycle was uniformly low under any NADP+/NADPH ratio. This is consistent with experimental observations using C$^{13}$ metabolic flux analysis by McCulley, \textit{et al.} \cite{McCulley2020} on \textit{R. rubrum} and McKinlay, \textit{et al.}. on \textit{R. palustris} \cite{McKinlay_2010}. Growth increased as thermodynamic odds moved from relatively more oxidative (<$10^{-4}$) toward relatively more reductive conditions (>$10^{-4}$), as did fatty acid synthesis. This is expected, because as the thermodynamic odds move toward 1.0, the available NADPH relative to NADP+ increases (Table \ref{tab:nad_nadph_odds}), and the free energy barrier for reduction of substrate to biomass decreases. Also as thermodynamic odds moved from relatively more oxidative toward relatively more reductive conditions, the primary routes for CO$_2$ assimilation were found to be the reactions of the reductive TCA cycle and pyruvate metabolism. 

Surprisingly, even under highly oxidative conditions (thermodynamic odds approaching $10^{-10}$) the TCA cycle did not operate in the fully oxidative direction.
Moreover, under all conditions examined, net NADPH consumption, rather than production, was the rule (\textit{vida infra}, Tables \ref{tab:overall_malate} and  \ref{tab:overall_acetate}). 
While the reverse reaction in Eqn \ref{NADPH_NADP_redox_reaction} is favored, the coupling of redox and anabolic reactions to ATP hydrolysis and other processes makes NADPH consumption more favorable. 

Although phototrophic growth under anaerobic conditions would naively be expected to result in a more reductive environment inside the cell where conversion of NADPH to NADP+ is favored  (reflective of thermodynamic odds approaching 1 on the right side of the x-axis of Figure \ref{fig:CO2vNADP}),
the simulation results under such conditions agree with inferences based on experimental observations made using C$^{13}$ metabolic flux analysis. This ostensibly implies that the internal redox environment \textit{in vivo} is relatively oxidized. Indeed, the simulation results are consistent with the C$^{13}$ metabolic flux analysis data only in the relatively oxidative range of NADP+/NADPH ratios from the center to left-hand side of the Figure  \ref{fig:CO2vNADP} x-axis (thermodynamic odds $\leq 10^{-7}$).  

There are of course caveats in both the model and the metabolic flux analysis that need to be considered. In metabolic flux analysis, a model is used to generate flux distributions which in turn are used to generate predicted isotope labelling patterns. The predicted isotope labeling patterns are then compared to experimentally measured isotope patterns, and the best-fitting computational isotope pattern implicates the analogous flux distribution. The inference can be very sensitive to the computational model that is used \cite{Theorell_2017}. The models used in the study by McCully, \textit{et al.} in \cite{McCulley2020} were understandably relatively small and also included the first few reactions of the oxidative pentose phosphate pathway \cite{McCulley2020}, which are not likely in the species. In addition, it was assumed that flux was unidirectional from isocitrate to 2-oxoglutarate for all models. In the simulation model, isocitrate flux was in the reductive direction in all cases. Thus, clarification of this issue is an open question.  

However, a direct comparison of the 250 reaction fluxes from the simulation model and the 64 reactions of the MFA model for malate of McCulley, \textit{et al.}. is difficult because there are only 18 reactions that overlap in the two models (not counting the malate uptake reaction). This is due to the limited size of the MFA model and because many of the 64 MFA reactions are composite reactions representing complex pathways of secondary metabolism that had no direct counterpart in the detailed simulation model. Of the 18 reactions that are in both models, at a thermodynamic odds of $10^{-7}$, 11 of these have flux in the same direction and these 10 reactions are all in central metabolism. Of the 8 reaction fluxes that do not agree, four are in the pentose phosphate pathway/Calvin-Benson-Bassham cycle, one is the isocitrate dehydrogenase reaction, one is in secondary metabolism (serine hydroxymethyltransferase), one is the malate synthase reaction, and one is the citrate synthase transformation. These reactions for each model are shown in Supplemental Table 1. 

For the comparison to the metabolic flux analysis model, we evaluated the thermo-kinetic models containing both the oxidative and reductive versions of the tricarboxylic acid (TCA) cycle, both seperately and together. 
Since the ATP-dependent citrate lyase, a key enzyme of the reductive TCA cycle, is missing from \textit{R. rubrum}, this has lead some to speculate that the reductive TCA cycle may not operate in \textit{rubrum} \cite{Beuscher1972}. Support for this hypothesis was the lack of detection of citrate and acetate in \textit{in vivo} assays even though earlier physiological and enzymatic evidence suggested a functioning, albeit low-flux reductive TCA cycle in \textit{R. rubrum} \cite{Buchanan1990, wang1993}. However, due to the transient nature of non-equilibrium concentrations, lack of observation cannot be used to draw conclusions. \textit{R. rubrum} does contain the ATP-independent citrate lyase and acetate CoA ligase enzymes, which together carry out the same transformation as ATP-dependent citrate lyase \cite{EISENBERG_1955,buchanan_1967}. Our model of the reductive TCA cycle includes the ATP-independent citrate synthase used in the oxidative TCA cycle as well as both  citrate lyase and acetate CoA ligase reactions.

Each version of the TCA cycle was complemented with the appropriate oxidative/reductive version of the pyruvate to acetyl-CoA reaction: either pyruvate synthase for the reductive TCA cycle or pyruvate dehydrogenase for the oxidative TCA cycle. 
The reductive and oxidative processes differ at both the reactions for converting pyruvate to acetyl-CoA and 2-oxoglutarate to succinyl CoA 
in which ferredoxin (Fd) is used for the reductive processes and NAD+ for the oxidative processes. The ferredoxin redox pairs are favored under reducing conditions since their reduction potential is slightly more favorable for carrying out reductive reactions,
\begin{equation}
2 Fd_{red} + NAD+ \rightleftharpoons 2 Fd_{ox} + NADH \textbf{\ \ \ \ } \Delta G = -15.5 \text{\ kJ/mol at pH\ }  7.0.  
\end{equation}
The results were not qualitatively different, in that in all three cases the TCA cycle ran similarly for each growth condition using both malate and acetate as the primary carbon source. The only quantitative difference was that the TCA cycle carried a slightly higher net flux when the reactions of the reductive TCA cycle were used under reductive conditions instead of the reactions of the oxidative TCA cycle. This illustrates a common misconception regarding cellular thermodynamics, that many reactions are not reversible unless a specific enzyme is present. Thermodynamics determines the net directionality of a reaction - always - and the role of the catalyst is to reduce the transition state barrier. Enzymes associated with specific directions of reactions, such as NAD-dependent 2-oxoglutarate dehydrogenase and ferredoxin-dependent 2-oxoglutarate synthase, are selected for by nature because they reduce thermodynamic costs for specific conditions. However, this does not mean that the reaction can go in only one direction. Given enough reactant, these seemingly one-way reactions can be reversed, as has been recently observed whereby the oxidative TCA cycle runs in reverse in the appropriate conditions \cite{Filipp_2012, Steffens_2021}. 
Since the TCA cycle mostly operated in the reductive direction across the range of redox conditions (Figure \ref{fig:CO2vNADP})  regardless of which enzymes were used, the results below pertain to the model using the reductive TCA cycle and the pyruvate ferredoxin oxidoreductase.

In support of the hypothesis that the internal environment is not highly reduced during growth on malate is the observation that the conditions in the model in which the flux through phosphoenolpyruvate carboxykinase (PEP-CK) best agrees with the experimental MFA data of McCulley, \textit{et al.} is when the thermodynamic odds of NADP+/NADPH is approximately $10^{-7}$ or below (Figure \ref{fig:CO2vNADP}). 
In both the experimental MFA fluxes and in the model under oxidative conditions during growth on malate, the reversible PEP-CK reaction had significant flux in the direction of conversion of oxaloacetate to phosphoenolpyruvate,
\begin{align*}
    \text{oxaloacetate + GTP} \rightleftharpoons \text{PEP + CO$_2$ + GDP,}    
\end{align*}
or alternately,
\begin{align*}
    \text{oxaloacetate + ATP} \rightleftharpoons \text{PEP + CO$_2$ + ADP.}    
\end{align*}
Under relatively more reductive conditions, the reaction in the model had no flux. 


In contrast, during growth on acetate the PEP-CK reaction had flux in the opposite direction, assimilating CO$_2$ and channeling carbon from phosphoenolpyruvate into oxaloacetate and then malate. The ratio of GTP:GDP and ATP:ADP remained relatively stable throughout the redox range, allowing the PE-PCK reaction to remain favorable for CO$_2$ assimilation.  
Consequently, CO$_2$ assimilation through PEP-CK does not decrease proportionately with a decrease in reductive power. Yet, significant reducing power is required to cycle the oxaloacetate produced by PEP-CK to malate and through the reductive TCA cycle into 2-oxoglutarate and pyruvate for biosynthetic purposes. Also interestingly during growth on acetate, the ethylmalony-CoA anaplerotic pathway did not produce malate and succinate under the simulation conditions. The ethylmalonyl-CoA pathway is thought to be essential for acetate growth for \textit{R. rubrum}. Unfortunately, previous metabolic flux analysis studies  for growth on acetate \cite{McKinlay_2010} were carried out before it was known that \textit{R. rubrum} did not contain the ethylmalonyl-CoA pathway and used the glyoxylate bypass instead. Therefore, lack of functionality of the ethylmalonyl-CoA in our simulations required further analysis.

Turning off the PEP-CK reaction alleviates the situation: CO$_2$ assimilation occurs primarily through the ethylmalonyl-CoA pathway during growth on acetate. Empirically, it has been found that high ATP concentrations, which would be expected during phototrophic growth, act to unidirectionally inhibit the PEP-CK reaction in the direction of oxaloacetate formation \cite{klemme_archMicrobiol_1976}. Consequently, in subsequent models used for acetate growth, the PEP-CK reaction was turned off. Using the ethylmalonyl-CoA pathway instead of the PEP-CK reaction to produce anapleroticaly made malate and succinate during growth on acetate may be favorable because the ethylmalonyl CoA pathway is more sensitive to redox conditions.
The overall reaction for the ethylmalonyl CoA pathway in \textit{R. rubrum} is,
\begin{align}
    \text{3 acetatyl-CoA + NADH + NADPH + FAD + CO$_2$ + H$_2$O} \rightleftharpoons \text{malate + propanoyl-CoA + NAD+ + NADP+ + FADH$_2$ + 2 CoA. }
    \label{ethylmalonylcoa_overall}
\end{align}
Two of the final steps in the pathway convert glyoxylate, water and acetyl-CoA to malate and CoA. This is carried out in two steps, with (S)-Malyl CoA as an intermediate. However, we observed that the single step malate synthase reaction, glyoxylate + acetyl CoA + H2O $\rightleftharpoons$ malate + CoA, can be substituted for the former with identical results. We were not able to utilize both sets of reactions (one set directly making malate and the other set making malyl-CoA as an intermediate from glyoxylate) at the same time, as doing so caused the ethylmalonyl-CoA pathway to go to equilibrium, stopping acetate assimilation. The model still grew when both sets of reactions were used, but by strictly CO$_2$ assimilation processes other than the ethylmalonyl-CoA pathway. 
Regarding the overall reaction of the ethylmalonyl-CoA pathway, that the pathway is favored under reductive conditions is especially apparent in that the lone oxidative component due to FAD is more than offset by NADPH in that the reaction FAD + NADPH $\rightleftharpoons$ FADH$_2$ + NADP+ favors the products by approximately 23 kJ/mol at pH 7.0.
The product propanoyl-CoA is then further metabolized to succinate with an overall reaction for the pathway,
\begin{align}
    \text{propanoyl-CoA + ATP + H$_2$CO$_3$} \rightleftharpoons \text{succinate + CoA + ADP + Pi}
    \label{ethylmalonylcoa_overall2}
\end{align}
We tested this for both acetate and malate growth by incrementing the standard free energy change for the ATP synthase reaction by $4\cdot RT \log 50$ such that ATP formation was driven by a 50-fold change in pH instead of a 10-fold change. Flux through the ethylmalonyl-CoA pathway roughly tripled, but never became the major pathway for acetyl-CoA assimilation compared to pyruvate synthase and the reverse TCA cycle. Interestingly, CO$_2$ assimilation through RubisCO and the CBB cycle increased modestly as well (not shown). 

Consistent with its role as an anaplerotic pathway, the ethylmalonyl-CoA pathway appears to function to provide substrates for the TCA cycle in relatively reductive conditions. In these conditions, acetyl-CoA cannot enter the TCA cycle in the oxidative direction because pyruvate is much more favorable than citrate formation. Consequently, some of the acetyl-CoA is converted to malate and succinate so that 2-oxoglutarate and oxaloacetate can be produced as precursors for amino acid synthesis. 

\hfill \break
\textbf{Biomass Oxidation State.} 
Finally, we compared the experimentally observed oxidation state of the biomass and ratio of produced macromolecules to the predicted oxidation of biomass from the simulation model. In the simulation model, the biomass is derived from the overall reaction stoichiometry for model growth where macromolecule ratios of DNA:RNA:protein:lipid produced are fixed in the ratio 1.0: 2.9: 44.1: 8.5 based on the original work of McCulley et. al. \cite{McCulley2020}. To demonstrate, under relatively oxidative conditions corresponding to a thermodynamic odds of NADP+:NADPH of $10^{-10}$ (Table \ref{tab:overall_malate}), the overall equation for growth on malate in the model  is,
\begin{align*}
   \text{10 C$_4$H$_6$O$_5$ + 3.8 NH$_3$ + 19.5 NADPH + 2.5 THF + 1.5 5,10-MTHF + 0.05 5-MTHF + 0.07 SO$_4$ + 2.2 ABP + 21.4 P$_i$} \\ 
   \rightarrow  \text{4.1 N10-fTHF + 3.1 P$_{i,2}$ + biomass  + 19.5 NADP+ + 4.5 CO$_2$} &
\end{align*}
in which 10 moles of malate (10 C$_4$H$_6$O$_5$) and 3.8 moles of NH$_3$ from the environment along with 19.5 moles of NADPH, 2.5 moles of tetrahydrofolate (THF), 1.5 moles of 5,10-methylenetetrahydrofolate, 0.05 moles of 5-methyl tetrahydrofolate, 0.07 moles of sulfate (SO$_4$), 2.2 moles of adenine-3,5-bisphosphate (ABP) and 21.4 moles of orthophosphate (P$_i$) are consumed to produce biomass and 4.1 moles of N10-formyltetrahydrofolate (N10-fTHF), 3.1 moles of diphosphate, 19.5 moles of NADP+ and 4.5 moles of CO$_2$.  
To obtain an estimate of the oxidation state of the biomass in the model, this reaction is simplified by removing the internal metabolites except the redox pair NADPH and NADP+, replacing biomass by a variable elemental composition, C$_4$H$_{u}$O$_{2.04}$N$_z$, and adding a variable amount of water to the right-hand side to give,
\begin{align*}
   \text{10 C$_4$H$_6$O$_5$ + 3.8 NH$_3$ + 19.5 NADPH }
   \rightarrow  x\ \text{C$_4$H$_{u}$O$_{2.04}$N$_z$ + $y$ H$_2$O + 4.5 CO$_2$} 
\end{align*}
For growth on malate, the macromolecular constraints such that DNA:RNA:protein:lipid is produced in the ratio 1.0: 2.9: 44.1: 8.5 (see methods) and the C:O value of 4:2.04 was used as it was supported by multiple measurements reported in Table \ref{tab:macro_ratios}. Use of a C:O value of 4:1.52 from \textit{R. palustris} reported by McCulley \cite{McCulley2020} increased discrepancy with simulation estimates.  The values of $x, y, z$ and $u$ are then obtained by balancing carbon, oxygen, nitrogen and hydrogen, respectively \cite{Doran2013}. The estimated biomass oxidation states for growth on malate under oxidative, neutral and reductive conditions are shown in Table \ref{tab:overall_malate}. Whether the redox conditions within the model are deemed to be oxidative, reductive or neutral is based, respectively, on whether CO$_2$ is produced, consumed, or neither produced nor consumed in significant quantities as before for the idealized model (Table \ref{tab:overall_malate}). The thermodynamic odds relating the NADPH concentrations to the NADP+ concentrations are shown for each condition as well. Regardless of whether CO$_2$ was produced or consumed, NADPH was consumed under all conditions, likely due to coupling of redox reactions to ATP hydrolysis during anabolism.

\input{malate_elemental_estimate}

\input{acetate_elemental_estimate}

As can be seen for the rows labeled \textit{oxidized} and \textit{neutral}, the estimated biomass elemental composition is consistent with values that have been observed experimentally, shown in Table \ref{tab:macro_ratios}, while the estimated biomass composition under reduced conditions has a high amount of hydrogen. These values are further evidence that \textit{in vivo}, the cell environment is relatively oxidizing despite the anaerobic photosynthetic conditions. 

For growth on acetate using the same model macromolecular constraints and C:O ratio, the estimated biomass composition ranged from C$_4$H$_{7.37}$O$_{1.57}$N$_{0.71}$ to C$_4$H$_{7.82}$O$_{1.57}$N$_{0.65}$, as shown in Table \ref{tab:overall_acetate}. In contrast to growth on malate, the hydrogen content of the biomass from growth on acetate is higher than the measured hydrogen content from the measured elemental composition of C$_4$H$_{6.94}$O$_{1.57}$N$_{0.63}$ \cite{Favier_2003}, indicating that, as one might expect, the growth substrate significantly affects the biomass composition and macromolecular ratios. And therefore for accurate thermo-kinetic modeling of metabolism, accurate biomass composition and macromolecular ratios are needed to constrain the model. 


Consequently, we measured the macromolecular ratios of DNA:RNA:protein:lipid for \textit{R. rubrum} grown on acetate. For comparison,  we also measured the macromolecular ratios produced when \textit{R. rubrum} is grown on malate, ethanol/CO$_2$, and butyrate/CO$_2$. 
The results are also shown in Table \ref{tab:macro_ratios}. First, the macromolecular ratios measured for growth on malate of 1.00: 2.96: 39.98: 9.40 are very consistent with those reported by McCully, \textit{et al.} at 1.0: 2.9: 44.1: 8.5. For growth on acetate, the measurements confirmed that \textit{R. rubrum} produces a significantly greater proportion of reduced compounds (lipids) when grown on acetate relative to malate. (Lipid has an elemental formula of approximately C$_4$H$_{7.4}$O$_{1.35}$.) The high lipid content is in sharp contrast to estimates of the redox state of biomass from elemental analysis in which the hydrogen content is not appreciably higher, C$_4$H$_{6.94}$O$_{1.57}$N$_{0.63}$. 
In fact, despite the high lipid production for cells grown on acetate, ethanol, and butyrate, the oxidation state of the biomass carbon is quantified by elemental analysis to be equal or more oxidized than that for malate-grown biomass.  

The reason for the predicted high hydrogen content of the acetate-grown biomass by the simulation versus measured by elemental analysis is likely two-fold. First, models are inherently incomplete representations of nature, and in this regard the inclusion of additional reductive pathways such as the polyhydroxybutyrate pathway and peptidoglycan pathway may bring the predicted elemental composition in closer agreement with the measured composition. Alternately, the model may be accurate and the discrepancy is rather due to the experimental sample preparation process. In preparation for elemental analysis, cells are heavily washed before lyophilization to remove media sulfate, phosphate, ammonia, and carbon growth substrate so they do not bias the measurement. During sample washing it is likely that many soluble, reduced hydrocarbons made by the cell are also washed away, likely effecting not only the C:H:O ratio, but also the C:N ratio. For growth on acetate, ethanol and butyrate, it may very well be that the actual C:H ratio are higher, in line with the macromolecular ratios and model predictions. 

\begin{table}[]
    \centering
    \begin{tabular}{|l|c|c||c|c|c|c||c|c|}
    \toprule
    & Measured & Carbon & \multicolumn{6}{c|}{Measured Macromolecular Ratios $^d$} \\
    Substrate	& Biomass & Ox. State & DNA	& RNA	& Protein &	Lipid & PHB  & Chromatophores \\
    \midrule
    Malate	& C$_4$H$_{7.20}$O$_{1.52}$N$_{0.72}$ $^a$ & -0.50 & 1.00	& 2.96	& 39.98	& 9.40 & 0.07 & 0.42 \\
    	& C$_4$H$_{7.56}$O$_{2.04}$N$_{0.56}$ $^b$ & -0.45 & & & & & & \\
    Acetate	& C$_4$H$_{6.94}$O$_{1.57}$N$_{0.63}$ $^c$ & -0.48 & 1.00 &	2.51 &	35.25 &	30.13 &	0.73 & 0.41 \\
    EtOH/CO$_2$ & C$_4$H$_{7.03}$O$_{2.03}$N$_{0.61}$ $^d$ & -0.29 &	1.00 & 2.98 & 36.59 & 18.38 & 4.20 & 0.37 \\
    Butyrate/CO$_2$ & C$_4$H$_{7.12}$O$_{2.05}$N$_{0.65}$ $^d$ & -0.27 & 1.00	& 2.44	& 31.32	& 9.06	& 9.66	& 0.42  \\
    \bottomrule
    \end{tabular}
    \caption{(Left) Experimentally measured elemental composition of cell biomass and estimated oxidation state of carbon from the elemental formula. (Right) Experimentally measured mass ratios of macromolecules relative to DNA. Elemental analysis sources: $^a$ elemental composition of \textit{Rhodopseudomonas palustris} biomass when grown on malate from Carlozzi and Sacchi \cite{carlozzi_2001}; $^b$ average elemental composition of \textit{Rhodobacter sphaeroides} from Walig\'{o}rska \textit{et. al} \cite{Waligorska_jAppMicrobiol_2009}, $^c$ average elemental composition of \textit{Rhodospirillum rubrum} growth on acetate from Favier-Teodorescu, \textit{et al.} \cite{Favier_2003}; $^d$ measurements for \textit{R. rubrum} in this study.}
    \label{tab:macro_ratios}
\end{table}


\section*{Discussion}

The question that we have sought to address is how the balance between oxidation and reduction is maintained during growth on malate and acetate. It is clear from the simulation results shown in Figure \ref{fig:CO2vNADP} and the macromolecular mass ratios shown in Table \ref{tab:macro_ratios} that the redox state of the cell can drive large changes in macromolecular synthetic pathways.

In comparing the simulation output with experimental results from the literature, it would appear that despite anaerobic phototrophic growth, the cellular environment is not very reductive. 

The experimental fluxes from MCA are consistent with NADP+/NADPH odds $< 10^{-7}$. In fact, the results even at the lower end of the tested range, an odds of 10$^{-10}$, are consistent with the experimental data.

These values of odds correspond to concentration ratios of NADP+/NADPH of approximately 7 to 7,000, respectively, while the respective ratios for NAD+/NADH concentrations are relatively consistent, in the range from 1000 to 10,000 for growth on both malate and acetate (Table \ref{tab:nad_nadph_odds}). Although typical experimentally measured values of these ratios vary roughly from 10 to 100, the predicted concentration ratios cannot be directly compared to experimental measurements of concentrations because experimental assays measure whole-cell concentrations, including both enzyme-bound and unbound concentrations, while the relevant thermodynamic values in the simulation correspond  to only the unbound species. It is the unbound concentrations that control the thermodynamics.
    
During photosynthesis, \textit{R. rubrum} can take advantage of reverse electron flow in the NADH-dehydrogenase reaction of the electron transport chain in which the proton motive force across the periplasmic membrane can be coupled to production of NADH,
    \begin{equation*}
        \text{NADH + Q + 5 H}_{cytoplasm}^+ \leftarrow \text{NAD+ + QH$_2$ + 4 H}_{periplasm}^+.
    \end{equation*}
    Although this process would aid production of NAD(P)H, it may do so at the expense of additional ATP production. Comparison of the reaction fluxes from the simulation model to inferences of reaction fluxes from isotope labeling data suggests that the internal environment remains relatively oxidized. Specifically, during growth on malate, if flux proceeds from malate to oxaloacetate and then to phosphoenolpyruvate via the PEP carboxykinase reaction as indicated by isotope labeling, then according to the simulation results the internal environment of the cell has to be fairly oxidative for this flux to be feasible.

Thus, rather than photosynthesis inducing high levels of NADH or NADPH, the proton gradient across the periplasmic membrane is likely to be used mostly for ATP production. 
    Yet, because of the driving force provided to anabolic reactions by ATP, reduction still occurs despite high values of NAD(P)+ relative to NAD(P)H.

In addition to adjusting fluxes of redox reactions in central metabolism \cite{McCulley2020} and using alternate assimilation pathways such as the ethylmalonyl CoA pathway \cite{Alber_B_2006}, the simulations suggest that a major route of maintaining redox balance is likely through dissipation of reducing equivalents by increased production of reduced metabolites such as fatty acids. However, this need not be limited to fatty acids.

\begin{table}[hbt]
\centering
\begin{tabular}{|cllllll|}
\toprule
\multicolumn{1}{|c}{Compound} & \multicolumn{1}{c}{Formula} & \multicolumn{1}{c}{C} & \multicolumn{1}{c}{H} & \multicolumn{1}{c}{O} & \multicolumn{1}{c}{N} & \multicolumn{1}{c|}{\textbf{S}} \\
\toprule
\textbf{Avg Nucleotide} & C$_{19}$H$_{20}$O$_{4}$N$_{15}$  & 1.74  & 1  & -2 & -3 & \\
\hline
\textbf{RNA} & C$_{38}$H$_{42}$O$_{20}$N$_{15}$ & 1.13 & 1  & -2  & -3 &\\
\hline
\textbf{DNA}            & C$_{39}$H$_{44}$O$_{16}$N$_{15}$   & 0.85  & 1  & -2 & -3 & \\
\hline
\textbf{Avg AA} \cite{senko1995}  & C$_{5}$H$_{7.7}$O$_{1.5}$N$_{1.4}$S$_{0.04}$\  & 0.05  & 1  & -2  & -3  & -0.2 \\
\hline
\textbf{PHB} & C$_4$H$_{6}$O$_{2}$ & -0.50 & 1 & -2 & &\\
\hline
\textbf{Lipid} & C$_4$H$_{7.4}$O$_{1.35}$ & -1.18  & 1 & -2 &   &\\
\bottomrule
\end{tabular}
\caption{Redox states of amino acids and nucleotides. Charge states of atoms were calculated with the python module OxidationNumberCalculator (https://github.com/Hiwen-STEM/OxidationNumberCalculator). Formula for Nucleotides, RNA, and DNA are for the average 4-mer A(T/U)GC.}
\label{tab:atomic_charge_states_summary}
\end{table}

Varying the relative levels of DNA, RNA, protein and lipids can impact the redox state of the cell. In support of this hypothesis is the fact that the average oxidation state of carbon in lipids, PHB and amino acids are much more reduced than the average oxidation state of carbon in RNA and DNA, as shown in Table \ref{tab:atomic_charge_states_summary} (see also Table \ref{tab:atomic_charge_states}, Supplementary Material). For example, the approximate reducing power to convert RNA to amino acids is given by the redox equation,
\begin{align*}
   \text{5 C$_{38}$H$_{42}$O$_{16}$N$_{15}$ + 98.9 NADPH }
   \rightarrow  38\ \text{C$_5$H$_{7.8}$O$_{1.5}$N$_{1.4}$ + 32.0 H$_2$O + 21.8 NH$_3$  + 98.9 NADP+.} 
\end{align*}
If we view, for example, nucleic acids and amino acids as an oxidation-reduction pair analogous to NAD+ and NADH, then we can rationalize that much of the reductant or high potential electrons that would normally require an electron acceptor such as CO$_2$, can likewise be accepted by nucleic acids and converted to amino acids, PHB or lipids. From this perspective, the relatively high protein mass fraction in the DNA:RNA:protein mass ratios of 1.0:2.9:44.1 found in the study by McCully, \textit{et. al} can be understood.

Biomass composition can vary considerably depending on the growth condition and species \cite{Beck_2018}. To test the hypothesis that the macromolecular ratios impact the redox state of the cell, we ran the model using DNA:RNA:protein:lipid ratios from two studies for different organisms of 1:6.6:17.7:8.5 \footnote{The value of 8.5 for lipid composition was estimated here.} (\textit{N. crassa}) \cite{talbot1982} and 1:17.2:35.2:6.7 (\textit{E. coli}) \cite{Beck_2018}. The estimated biomass compositions are shown in Table \ref{tab:overall_acetate2}; the results confirm the hypothesis that varying the ratios of DNA:RNA:protein:fatty acid can significantly impact the redox state. In fact, varying the macromolecule ratios can have a larger impact on biomass redox state than varying the NADPH:NADP+ odds, as can be seen in comparing the middle and top rows of Table \ref{tab:overall_acetate}, in which the NADP+:NADPH ratio changes by 3 orders of magnitude, to the results shown in Table \ref{tab:overall_acetate2}, in which the DNA:RNA:protein:fatty acid ratio changes from 1.0: 2.9: 44.1: 8.5 to 
1: 6.6 : 17.7 : 8.5 (\textit{N. crassa}) and
1: 17.2: 35.2: 6.7(\textit{E. coli}).
\input{acetate_alternate_cellComposition.tex}

Furthermore, since DNA is produced only during part of the cell cycle, it is apparent that the redox state of a cell would also have an impact on the dynamics of the cell cycle, and likewise, the cell cycle would have an impact on the redox state of the cell.
It is likely that the redox state of the cell undergoes cycling in sync with the bacterial cell cycle. Such cycling may demonstrate how biological cells, as dissipative structures, use dissipation to drive memory mechanisms, the DNA processes and dynamics, that not only allows the cell to predict future events from past experience, but also record memories of new environmental patterns as well. 

\section*{Author Contributions}

WRC constructed the metabolic model with help from JAN. WRC developed the thermodynamic methods describing the mass action kinetics. EK developed the control and optimization method to constrain metabolite concentrations and optimize growth. WRC carried out all optimizations and analyzed the data. KAH and JAN performed the experimental quantification of macromolecules. WRC, JAN and EK wrote the article. 

\section*{Acknowledgments}

WRC and JAN were supported by the U.S. Department of Energy, Office of Biological and Environmental Research through contracts 78266 and award DE-SC0022091, respectively. EK was supported by the Data Model Convergence Initiative and by the Predictive Phenomics Initiative, under the Laboratory Directed Research and Development Program at at the Pacific Northwest National Laboratory. PNNL is a multiprogram national laboratory operated by Battelle for the U.S. Department of Energy under contract DE-AC05-76RLO 1830. The authors are grateful to Tom Wietsma for providing the \textit{R. ruburm} elemental composition measurements at the PNNL Environmental Molecular Sciences Laboratory supported by award 60175 to JAN and WRC.

\section*{Supplementary Material}

An online supplement to this article can be found by visiting BJ Online at \url{http://www.biophysj.org}.


\bibliography{biblio2}

\begin{figure}[hbt]
    \centering
    \includegraphics[width=0.9\linewidth]{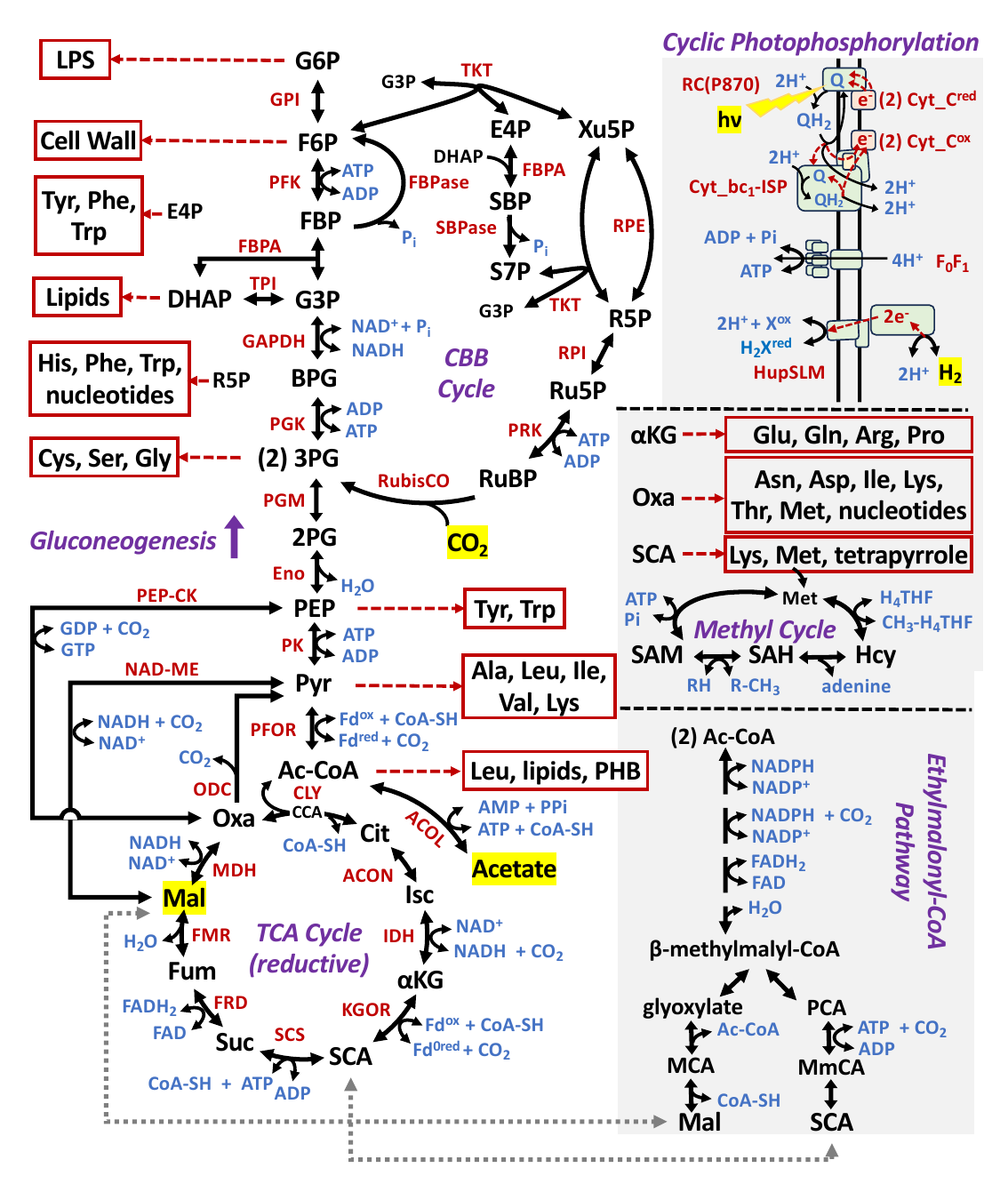}
\end{figure}
\begin{figure}
    \contcaption{Overview of cyclic photophosphorylation by the branched-cyclic Q cycle, hydrogen uptake, and primary and secondary biosynthetic pathways of \textit{Rhodospirillum rubrum} included in the thermo-kinetic model. Red dashed arrows indicate flow of single electrons in the Q cycle unless otherwise noted, and each cytochrome C (Cyt\_C) carries 1-e$^-$ each, so two (2) Cyt\_C proteins are required. For hydrogen uptake, note the specific redox carrier that interacts with the HupSLM [Ni-Fe] hydrogenase is not known, and thus indicated by X/H$_2$X for a general thermodyanmically acceptable redox carrier, e.g. NAD(P)+/NAD(P)H.  Both the oxidative (not shown) and reductive (shown) versions of the TCA cycle were tested in the modeling. The reductive TCA cycle was used in the final model. \textbf{Metabolites} (alphabetical order): 2PG - 2-phospho-glycerate, 3PG - 3-phospho-glycerate (2 indicates the formation of 2 molecules from RuBP), Ac-CoA - acetyl-CoA, $\alpha$ KG - 2-keto-glutarate, BPG - 1,3-bisphospho-glycerate, Cit - citrate, CCA - citryl-CoA (intermediate of prokaryotic ATP-independent citrate lyase, CYL), CoA-SH - Coenzyme A, DHAP - dihydroxyacetone phosphate, E4P - erythrose-4-phosphate, F6P - fructose-6-phosphate, Fum - fumarate, G3P - glyceraldehyde-3-phosphate, G6P - glucose-6-phosphate, H4THF – Tetrahydrofolate, Hcy – homocysteine, Isc - isocitrate, LPS - lipopolysaccharide, Mal - malate, MCA - malyl-CoA, MmCA - methylmalyl-CoA, Oxa - oxaloacetate, PCA - propanoyl-CoA, PEP - phosphoenol-pyruvate, PHB - polyhydroxybutyrate, Pyr - pyruvate, R-CH3 - methylated methyl acceptor (R), Q/QH$_2$ - oxidized and 2e$^-$ reduced quinone, R5P - ribose-5-phosphate, Ru5P - ribulose-5-phosphate, RuBP - ribulose-1,5-bisphosphate, SAM – S-adenosyl-methionine, SAH – S-adenosyl-homocysteine, S7P - sedoheptulose-7-phosphate, SBP - sedoheptulose-1,7-bisphosphate, SCA - succinyl-CoA, Suc - succinate, Xu5P - xylulose-5-phosphate,  X$^{ox}$/H$_2$X$^{red}$ - a redox carrier. All amino acids are indicated by standard 3-letter code.
    \textbf{Enzymes} (alphabetical order): ACOL - acetate:CoA ligase (AMP-forming), ACON – aconitase, CLY – ATP-independent citrate lyase (forms citryl-CoA as an intermediate), Cyt\_bc1-ISP - complex of cytchrome b (heme b560 and b566 containing) cytochrome c1 (heme c containing) and Rieske iron sulfur protein (ISP, yellow), Cyt\_C - bacterial cytochrome C2 (heme c containing), Eno – enolase, F$_0$F$_1$ - ATP synthasae, FBPA – fructose-1,6-bisphospate aldolase, FBPase – fructose-1,6-bisphosphatase phosphatase, FRD – fumarate reductase, FUM – fumarase, GAPDH – glyceraldehyde-3-phosphate dehydrogenase, GPI – glucose-6-phosphate isomerase, HupSLM - Uptake hydrogenase complex of small, large, and medium subunits, IDH – isocitrate dehydrogenase, KGOR – 2-keto-glutarate:ferredoxin oxidoreductase ($\alpha$-KG synthase), MDH – malate dehydrogenase, NAD-ME – NAD-dependent malic enzyme, ODC – oxaloacetate decarboxylase, PEP-CK – phosphoenol-pyruvate  carboxykinase, PFK – phosphofructokinase, PFOR – pyruvate-ferredoxin oxidoreductase (pyruvate synthase), PGK – 3-phosphoglycerate kinase, PGM – 3-phosphoglycerate mutase, PK – pyruvate kinase, PRK – phosphoribulokinase, RC(P870) - Type II photosynthetic reaction center with characteristic P870 pigment for photo-oxidation of Cyt\_C$^{red}$, RPE – ribulose-5-phosphate epimerase, RPI – ribose-5-phosphate isomerase, RubisCO – ribulose-1,5-bisphosphate oxygenase/carboxylase, SBPase – sedoheptulose-1,7-bisphosphate phosphatase, SCS – succinyl-CoA synthetase, TKT – transketolase, TPI – triose-phosphate isomerase.} 
    \label{fig:pathway_overview}

\end{figure}

\begin{figure}[hbt!]
\centering
\includegraphics[width=0.65\linewidth]{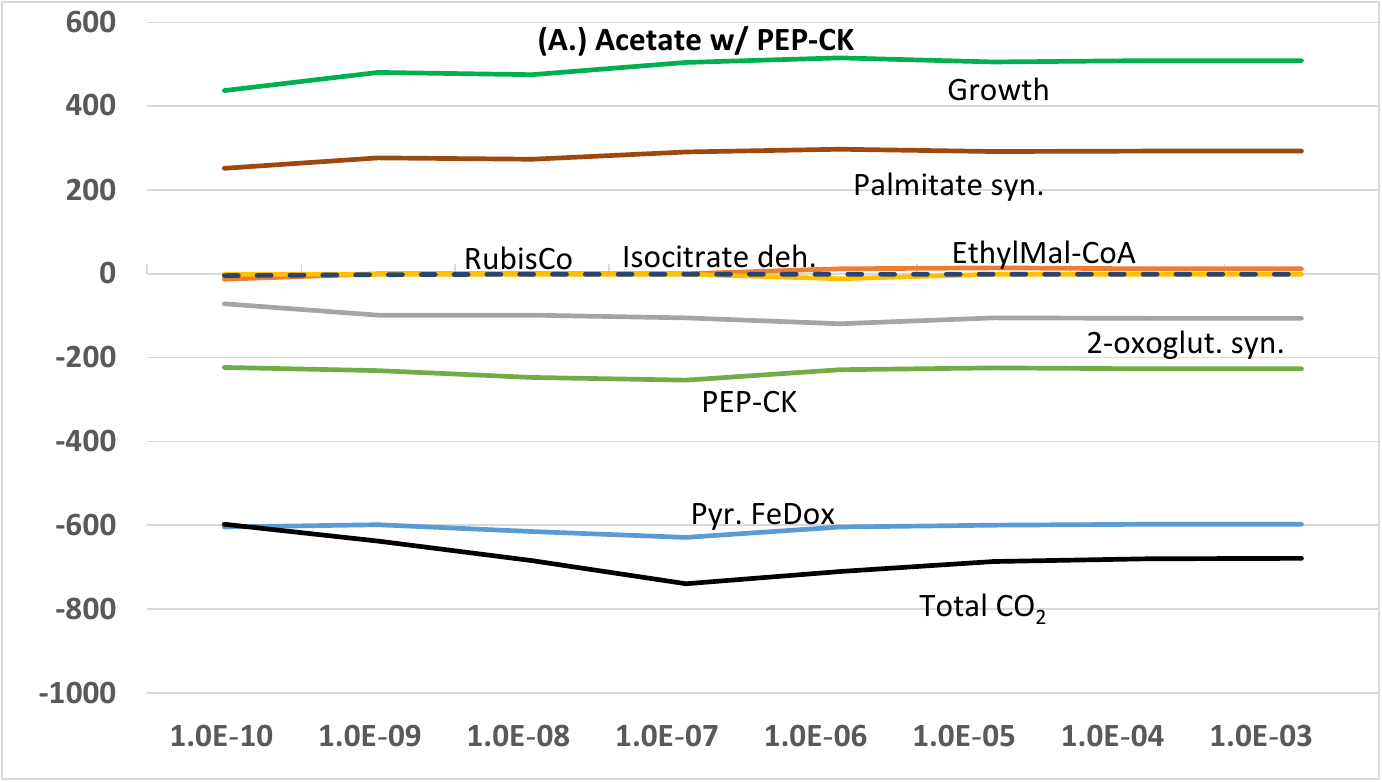}
\includegraphics[width=0.65\linewidth]{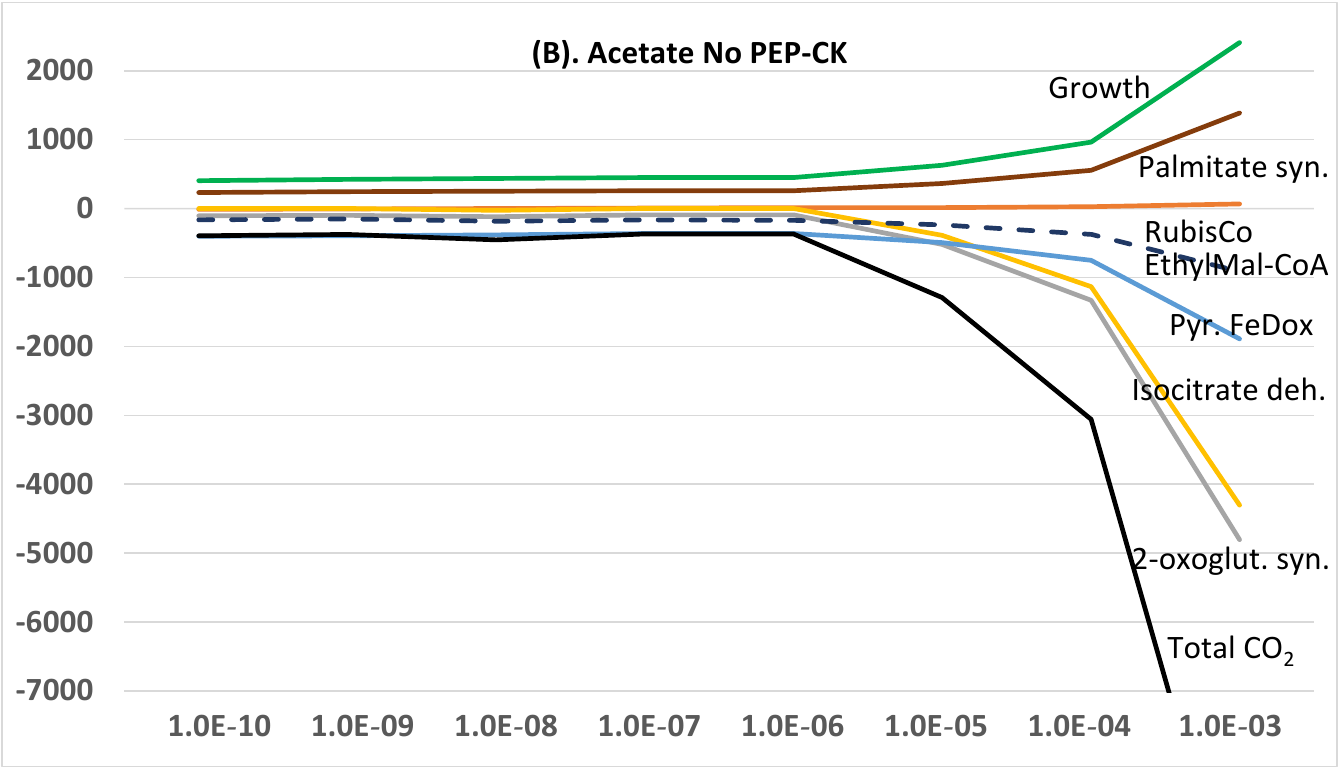}
\includegraphics[width=0.65\linewidth]{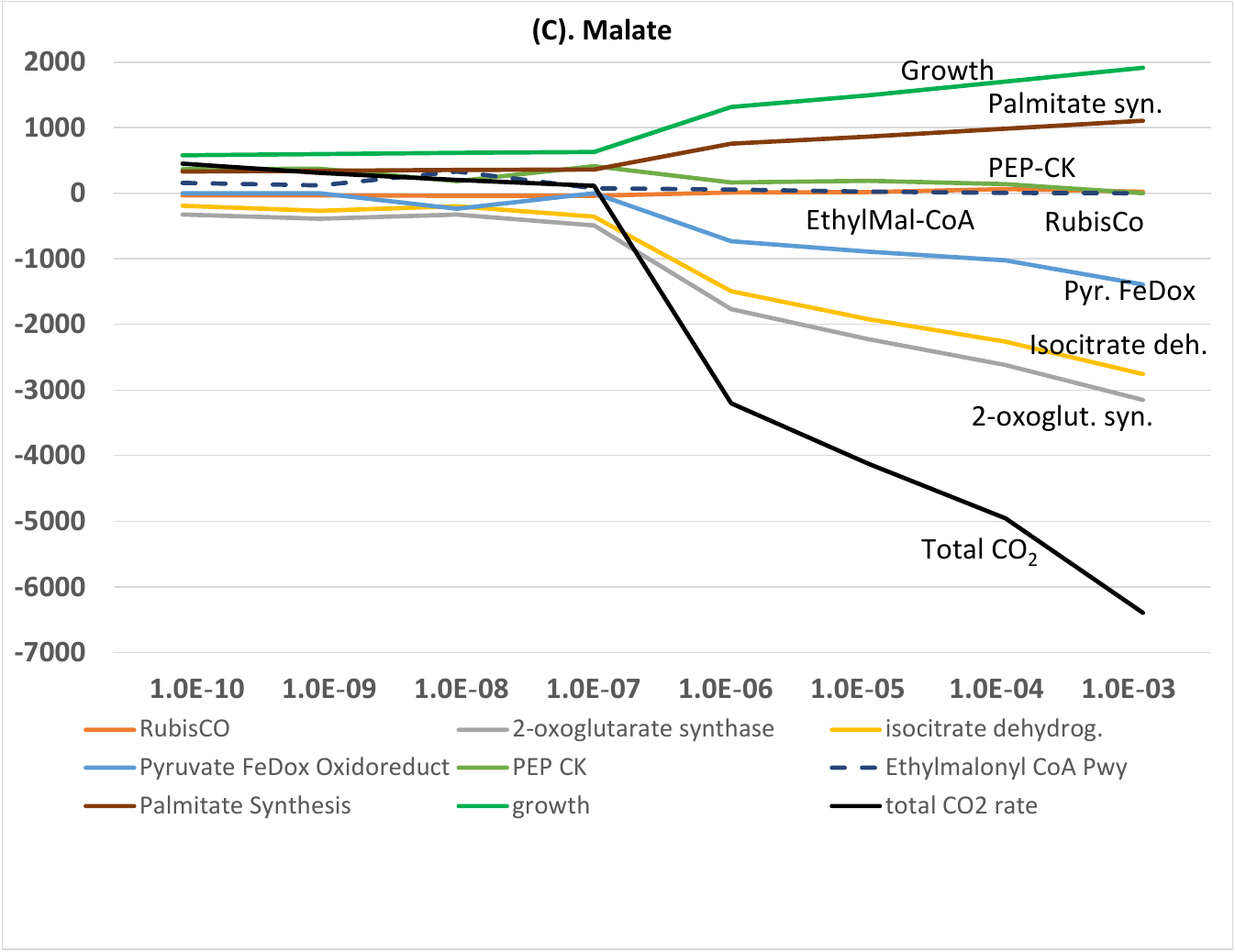}
\caption{Plot growth rate, palmitate (fatty acid synthesis), and rates of CO$_2$ flux at RubisCO Ethylmalonyl CoA pathways, 2-oxoglutarate synthase, isocitrate dehydrogenase, pyruvate ferredoxin oxidoreductase, phosphoenolpyruvate carboxykinase (PEPCK) as a function of the thermodynamic odds (KQ$^{-1}$) of  NADP+:NADPH. A negative vlaue indicates CO$_2$ assimilation and a positive value indicates CO$_2$ release. A thermodynaic odds (KQ$^{-1}$, Eqn \ref{eq:thermodynamic_odds}) of 10$^{-3}$ indicates that ratio of products to reactants (Q) in the chemical reaction NADPH $\rightleftharpoons$ NADP+ + H$_2$  is 1000-fold higher compared to the equilibrium ratio (K).}
\label{fig:CO2vNADP}
\end{figure}

\clearpage

\clearpage

\include{supplemental_information}

\end{document}

%% file: malate_elemental_estimate.tex
\definecolor{light-gray}{gray}{0.9}

\bgroup
\def\arraystretch{1.5}
\begin{table}[hbt]
    \centering
    \small
    \begin{tabular}{l|l|rcl}
\toprule
& Eqn \ref{eq:thermodynamic_odds}: & \\
State & Odds &  \multicolumn{3}{c}{Overall Reaction}  \\
\midrule
 &  & \cellcolor{light-gray}2.52 THF + 1.49 \ 5,10-MTHF + 0.05 5-MTHF + \\
 oxidized & $10^{-10}$ & \cellcolor{light-gray} 0.07 SO$_4$ + 2.18 ABP + 21.37 P$_i$ + & $\rightarrow$ & \cellcolor{light-gray} 4.06 N10-fTHF + 3.08 P$_{i,2}$ + \\ 
 & & 10 C$_4$H$_6$O$_5$ + 3.76 NH$_3$ + 19.46 NADPH  & &  \ \ \ \ \ \ \ biomass \ \ \ + \ \ \ \ 19.46 NADP+ + 4.52 CO$_2$\\
\hline
C Ox. State & -0.48 & 10 C$_4$H$_6$O$_5$ + 3.76 NH$_3$ + 19.46 H$_2$\ \ \ \ \ \ \ \ \ \   
 & $\rightarrow$ & 8.87 \textbf{C$_4$H$_{7.27}$O$_{2.04}$N$_{0.42}$} + 22.87 H$_2$O + 4.52 CO$_2$\\
\hline \hline
 &  & \cellcolor{light-gray}2.2 THF + 1.0 \ 5,10-MTHF + 0.06 5-MTHF + \\
neutral & $10^{-6.99}$ & \cellcolor{light-gray} 0.08 SO$_4$ + 1.85 ABP + 18.5 P$_i$ + & $\rightarrow$ & \cellcolor{light-gray} 3.3 N10-fTHF + 4.06 P$_{i,2}$ + \\ 
 &  & 10 C$_4$H$_6$O$_5$ + 5.32 NH$_3$ + 30.7 NADPH + 0.1 CO$_2$  & &  biomass + 30.7 NADP+ \\
\hline
C Ox. State & -0.50 & 10 C$_4$H$_6$O$_5$ + 5.32 NH$_3$ + 30.7 H$_2$\ \ \ \ \ \ \ \ \ \ + 0.1 CO$_2$  
& $\rightarrow$ & 10 \textbf{C$_4$H$_{7.7}$O$_{2.04}$N$_{0.53}$} + 29.6 H$_2$O  \\
\hline \hline
 &  & \cellcolor{light-gray}1.43 THF + 0.79 \ 5,10-MTHF + 0.18 5-MTHF + \\
reduced & $10^{-3}$ & \cellcolor{light-gray} 0.23 SO$_4$ + 1.71 ABP + 9.57 P$_i$ + & $\rightarrow$ & \cellcolor{light-gray} 2.4 N10-fTHF + 3.83 P$_{i,2}$ + \\ 
  & & 10 C$_4$H$_6$O$_5$ + 20.50 NH$_3$ + 187.2 NADPH + 67.61 CO$_2$  & &  biomass + 187.2 NADP+ \\
\hline
C Ox. State & -0.60 & 10 C$_4$H$_6$O$_5$ + 20.50 NH$_3$ + 187.2 H$_2$\ \ \ \ \ \ \ \ \ \ + 67.61 CO$_2$  
& $\rightarrow$ & 26.9 \textbf{C$_4$H$_{8.74}$O$_{2.04}$N$_{0.76}$} + 130.34 H$_2$O  \\
\bottomrule
\end{tabular}
\caption{Elemental biomass composition (\textbf{bold}) estimated from overall chemical reactions for growth on malate. Estimates were obtained from simulations under oxidizing, neutral and reducing redox conditions, corresponding to the concentrations of NADP:NADPH being held at the thermodynamic odds (Eqn \ref{eq:thermodynamic_odds}) of 10$^{10}$, $9.9 \times 10^6$ and 10$^{3}$, respectively. The rows labeled \textit{oxidized, neutral} or \textit{reduced} are those from the optimization for growth of the metabolic model while those labeled estimate are estimates of the biomass elemental composition using the stoichiometry found in the model. The grey rows are for those compounds in the overall chemical equation for the metabolic model that are not used in estimating the elemental composition of the biomass. \textbf{Abbreviations:} ABP: adenine-3,5-bisphosphate; THF: tetrahydrofolate; 5,10-MTHF: 5, 10 methylenetetrahydrofolate; N10-fTHF: N10-formyltetrahydrofolate; P$_{i,2}$: diphosphate.}
\label{tab:overall_malate}
\end{table}
\egroup

%% file: acetate_elemental_estimate.tex
\bgroup
\def\arraystretch{1.5}
\begin{table}[hbt]
    \centering
    \small
    \begin{tabular}{l|l|rcl}
\toprule
& Eqn \ref{eq:thermodynamic_odds} & \\
State & Odds & \multicolumn{3}{c}{Overall Reaction}  \\
\midrule
oxidized & $10^{-10}$& \cellcolor{light-gray}2.5 THF + 1.5 \ 5,10-MTHF + 0.05 5-MTHF + \\
  & & \cellcolor{light-gray} 0.07 SO$_4$ + 2.2 ABP + 20.6 P$_i$ + & $\rightarrow$ & \cellcolor{light-gray} 4.0 N10-fTHF + 3.1 P$_{i,2}$ + \\ 
 & & 10 C$_2$H$_4$O$_2$ + 4.23 NH$_3$ + 14.20 NADPH + 3.95 CO$_2$ & &  \ \ \ \ \ \ \ biomass \ \ \ + \ \ \ \ \ \ \ 14.20 NADP+ \\
\hline
C Ox. State & -0.52 & 10 C$_2$H$_4$O$_2$ + 4.23 NH$_3$ + 14.20 H$_2$\ \ \ \ \ \ \ \ \ \ + 3.95 CO$_2$   
& $\rightarrow$ &  5.99 \textbf{C$_4$H$_{7.37}$O$_{1.57}$N$_{0.71}$} + 18.48 H$_2$O \\
\hline \hline
neutral & $10^{-7}$& \cellcolor{light-gray}0.85 THF + 0.05 \ 5,10-MTHF + 0.04 5-MTHF + \\
& & \cellcolor{light-gray} 0.05 SO$_4$ + 0.59 ABP + 21.06 P$_i$ + & $\rightarrow$ & \cellcolor{light-gray} 0.95 N10-fTHF + 10.37 P$_{i,2}$ + \\ 
 & & 10 C$_2$H$_4$O$_2$ + 3.81 NH$_3$ + 15.26 NADPH + 3.65 CO$_2$  & &  biomass + 15.26 NADP+ \\
\hline
C Ox. State & -0.68 & 10 C$_2$H$_4$O$_2$ + 3.81 NH$_3$ +  15.26 H$_2$\ \ \ \ \ \ \ \ \ \ + 3.65 CO$_2$  
& $\rightarrow$ & 5.91 \textbf{C$_4$H$_{7.77}$O$_{1.57}$N$_{0.64}$} + 18.00 H$_2$O  \\
\hline \hline
reduced & $10^{-3}$& \cellcolor{light-gray}4.51 THF + 0.09 \ 5,10-MTHF + 0.22 5-MTHF + \\
& & \cellcolor{light-gray} 0.29 SO$_4$ + 3.05 ABP + 19.2 P$_i$ + & $\rightarrow$ & \cellcolor{light-gray} 4.83 N10-fTHF + 11.90 P$_{i,2}$ + \\ 
 & & 10 C$_2$H$_4$O$_2$ + 20.5 NH$_3$ + 255.16 NADPH + 106.11 CO$_2$  & &  biomass + 255.16 NADP+ \\
\hline
C Ox. State & -0.68 & 10 C$_2$H$_4$O$_2$ + 20.5 NH$_3$ + 255.16 H$_2$\ \ \ \ \ \ \ \ \ \ + 106.11 CO$_2$  
& $\rightarrow$ &  31.53 \textbf{C$_4$H$_{7.82}$O$_{1.57}$N$_{0.65}$} + 182.66 H$_2$O  \\
\bottomrule
\end{tabular}
\caption{Elemental biomass composition (\textbf{bold}) estimated from overall chemical reactions for growth on acetate. Estimates were obtained from simulations under oxidizing, neutral and reducing redox conditions, corresponding to the concentrations of NADP:NADPH being held at the thermodynamic odds (Eqn \ref{eq:thermodynamic_odds}) of 10$^{10}$, $10^7$ and 10$^{3}$, respectively. The rows labeled \textit{oxidized, neutral} or \textit{reduced} are those from the optimization for growth of the metabolic model while those labeled estimate are estimates of the biomass elemental composition using the stoichiometry found in the model. The grey rows are for those compounds in the overall chemical equation for the metabolic model that are not used in estimating the elemental composition of the biomass. \textbf{Abbreviations:} ABP: adenine-3,5-bisphosphate; THF: tetrahydrofolate; 5,10-MTHF: 5, 10 methylenetetrahydrofolate; N10-fTHF: N10-formyltetrahydrofolate; P$_{i,2}$: diphosphate.}
\label{tab:overall_acetate}
\end{table}
\egroup

%% file: acetate_alternate_cellComposition.tex
\bgroup
\def\arraystretch{1.5}
\begin{table}[hbt]
    \centering
    \small
    \begin{tabular}{l|l|rcl}
\toprule
DNA:RNA: & Eqn \ref{eq:thermodynamic_odds}\\
protein: lipid & Odds & \multicolumn{3}{c}{Overall Reaction}  \\
\midrule
& & \cellcolor{light-gray} 0.54 THF + 0.03 \ 5,10-MTHF + 0.03 5-MTHF + \\
\textit{N. crassa} & $10^{-7}$ & \cellcolor{light-gray} 0.03 SO$_4$ + 0.56 ABP + 22.49 P$_i$ + & $\rightarrow$ & \cellcolor{light-gray} 0.60 N10-fTHF + 10.56 P$_{i,2}$ + \\ 
1: 6.6: 17.7: 8.5 & & 10 C$_2$H$_4$O$_2$ + 3.09 NH$_3$ + 16.23 NADPH + 3.29 CO$_2$ & &  biomass + 16.23 NADP+ \\
\hline
esitmate & & 10 C$_2$H$_4$O$_2$ + 3.09 NH$_3$ + 16.23 H$_2$\ \ \ \ \ \ \ \ \ \ + 3.29 CO$_2$   
& $\rightarrow$ &  5.82 \textbf{C$_4$H$_{8.98}$O$_{2.04}$N$_{0.53}$} + 14.70 H$_2$O \\
\hline \hline

& & \cellcolor{light-gray} 0.94 THF + 0.18 \ 5,10-MTHF + 0.04 5-MTHF + \\
\textit{E. coli} & $10^{-7}$ & \cellcolor{light-gray} 0.05 SO$_4$ + 1.01 ABP + 22.62 P$_i$ + & $\rightarrow$ & \cellcolor{light-gray} 1.16 N10-fTHF + 10.70 P$_{i,2}$ + \\ 
1: 17.2: 35.2: 6.7 & & 10 C$_2$H$_4$O$_2$ + 3.77 NH$_3$ + 15.35 NADPH + 4.22 CO$_2$  & &  biomass + 15.35 NADP+ \\
\hline
esitmate & & 10 C$_2$H$_4$O$_2$ + 3.77 NH$_3$ +  15.35 H$_2$\ \ \ \ \ \ \ \ \ \ + 4.22 CO$_2$  
 & $\rightarrow$ & 10 \textbf{C$_4$H$_{7.95}$O$_{2.04}$N$_{0.62}$} \ \ + \ \ 16.10 H$_2$O  \\
\bottomrule
\end{tabular}
\caption{Elemental biomass composition (\textbf{bold}) estimated from overall chemical reactions for growth on acetate under varying levels of DNA:RNA:protein (P):fatty acid (FA). Estimates were obtained from simulations under redox conditions corresponding to the concentrations of NADP+:NADPH being held at the thermodynamic odds (Eqn \ref{eq:thermodynamic_odds}) of 10$^{7}$. The rows labeled with DNA:RNA:protein:lipid values are those from the optimization for growth of the metabolic model while those labeled estimate are estimates of the biomass elemental composition using the stoichiometry found in the model, and can be compared to the similar condition shown for acetate growth shown in Table \ref{tab:overall_acetate} in which the respective levels were 1.0: 2.9: 44.1: 8.5. The grey rows are for those compounds in the overall chemical equation for the metabolic model that are not used in estimating the elemental composition of the biomass. \textbf{Abbreviations:} ABP: adenine-3,5-bisphosphate; THF: tetrahydrofolate; 5,10-MTHF: 5, 10 methylenetetrahydrofolate; N10-fTHF: N10-formyltetrahydrofolate; P$_{i,2}$: diphosphate.}
\label{tab:overall_acetate2}
\end{table}
\egroup

%% file: supplemental_information.tex
\section*{Supplementary Information}
\subsection*{Models and Code}
The metabolic model was implemented as described in \textit{Methods} and is available in a python-based Jupyter notebook at DOI:10.5281/zenodo.10475680.

\subsection*{Fixed Metabolites}

\begin{table}[h]
\begin{tabular}{r|r}
\toprule
\multicolumn{1}{c}{Fixed Metabolite}               & \multicolumn{1}{c}{Concentration} \\
\toprule
(4S)-4,5-DIHYDROXYPENTAN-2,3-DIONE:CYTOPLASM & 2.00 $\times 10^{-09}$                          \\
5-METHYLTETRAHYDROFOLATE:CYTOPLASM           & 2.00 $\times 10^{-03}$                          \\
5,10-METHYLENETETRAHYDROFOLATE:CYTOPLASM     & 2.00 $\times 10^{-03}$                          \\
A   7,8-DIHYDROFOLATE:CYTOPLASM              & 2.00 $\times 10^{-03}$                          \\
ACYL-CARRIER   PROTEIN:CYTOPLASM             & 1.00 $\times 10^{-04}$                          \\
ADENOSINE-3',5'-BISPHOSPHATE:CYTOPLASM       & 1.00 $\times 10^{-03}$                          \\
AN OXIDIZED   C-TYPE CYTOCHROME:EXTERNAL     & 5.85 $\times 10^{-26}$                          \\
AN\_OXIDIZED\_FERREDOXIN:CYTOPLASM           & 2.00 $\times 10^{-03}$                          \\
AN\_OXIDIZED\_THIOREDOXIN:CYTOPLASM          & 1.00 $\times 10^{-04}$                          \\
CELL                                         & 1.00 $\times 10^{-09}$                          \\
CO2:CYTOPLASM                                & 1.00 $\times 10^{-04}$                          \\
COA:CYTOPLASM                                & 1.40 $\times 10^{-06}$                          \\
DIPHOSPHATE:CYTOPLASM                        & 1.00 $\times 10^{-04}$                          \\
FADH2:CYTOPLASM                              & 2.00 $\times 10^{-03}$                          \\
H2O:CYTOPLASM                                & 5.55 $\times 10^{+01}$                          \\
N10-FORMYLTETRAHYDROFOLATE:CYTOPLASM         & 2.00 $\times 10^{-03}$                          \\
NAD+:CYTOPLASM                               & 2.10 $\times 10^{-04}$                          \\
NADP+:CYTOPLASM                              & 8.86 $\times 10^{-06}$                          \\
NADPH:CYTOPLASM                              & 1.20 $\times 10^{-06}$                          \\
NH3:CYTOPLASM                                & 1.00 $\times 10^{-04}$                          \\
ORTHOPHOSPHATE:CYTOPLASM                     & 2.00 $\times 10^{-02}$                          \\
PHOTON:CYTOPLASM                             & 1.00 $\times 10^{-04}$                          \\
SULFATE:CYTOPLASM                            & 1.00 $\times 10^{-04}$                          \\
TETRAHYDROFOLATE:CYTOPLASM                   & 2.00 $\times 10^{-03}$                          \\
UBIQUINOL:CYTOPLASM                          & 2.00 $\times 10^{-03}$  \\
\bottomrule
\end{tabular}
\caption{The metabolites listed were held fixed as Dirichlet boundary conditions during all steady state optimizations. In addition, either malate or acetate, depending on the growth conditions was also held fixed at a concentration of 10 mM.}
\label{tab:fixed_metabolites}
\end{table}

\subsection*{Calculation of Reference Free Energies of Reaction}
Reference free energies of reaction are taken to be the standard free energies of reaction adjusted for an aqueous solution of ionic strength of 0.15 M and a pH of 7.0.
When standard free energy of reaction values cannot be found in eQuilibrator, the reaction free energies were estimated by various schemes listed below.

\begin{outline}
\1 Reference free energies for all reactions involving an exchange of indistinguishable metabolites between compartments were assumed to zero. 
    
\1 Values for the chemical potential for S-adenosyl-L-methionine (SAM) could not be calculated. Consequently, the reference free energies of reaction for the two reactions involving SAM were taken to be zero:
\begin{eqnarray*}
\text{S-ADENMETSYN-RXN:     ATP + L-methionine + H2O}  & \longrightarrow & \text{SAM + phosphate + diphosphate}   \\
\text{RXN-7605:\ \ \ \ \ \ \ \ SAM + tetrahydrofolate} & \longrightarrow & \text{S-adenosyl-L-homocysteine + 5-methyltetrahydrofolate}     
\end{eqnarray*}

\1 Reference free energies could not be calculated for,
\begin{eqnarray*}
    \text{RXN-15121: \ \ \ \ \ \ \ \ } \text{2-aminobut-2-enoate} & \longrightarrow & \text{2-iminobutanoate} \\
\text{RXN-15123}: \text{2-iminobutanoate + H2O} & \longrightarrow & \text{2-oxobutanoate + NH3}
\end{eqnarray*}
However, the combined reaction,
\begin{equation*}    
 \text{2-aminobut-2-enoate + H2O} \longrightarrow \text{2-oxobutanoate + NH3} 
\end{equation*}
had a reference free energy of -34 kJ/mol. Consequently, each reaction was set to have a reference free energy of half the value (-17 kJ/mol) of the combined reaction.

\1 Reactions that include cytochromes,
\begin{eqnarray*}
    \text{RXN1ZKB-9: 2 photons + quinone + cyt-c(2+)} & \longrightarrow & \text{quinol + 2 cyt-c(3+)} \\
    \text{1.10.2.2-RXN: \ \ \ \ \ \ \ \ \ \ \ \ \ \ \ \ \ \ \ \  quinol + 2 cyt-c(3+)} & \longrightarrow & \text{quinone + cyt-c(2+) 4 H+(periplasm)} 
\end{eqnarray*}
were calculated according to the following electrochemical scheme,
\begin{table}[h!]
    \centering
    \begin{tabular}{ccccl}
 2 Cyt-c(2+) & $\longrightarrow$ & 2 Cyt-c(3+) + 2e-   & 2* $E^\circ$ &= ~ 0.293V \\ 
 Ubiquinone + 2 e- & $\longrightarrow$ & Ubiquinol  & 2 * $E^{\circ}$ &= -0.083 V \\
 \hline
 2 Cyt-c(2+) + Ubiquinone & $\longrightarrow$ & 2 Cyt-c(3+) + Ubiquinol & $\Delta G^{\circ}$ &= -z$F\Delta$ E0 \\
 & & & & = -2(96.485kJ/V)(0.210V) \\
 & & & & = -40.52 kJ/mol 
    \end{tabular}
    \label{tab:my_label}
\end{table}
\\
such that reactions RXN1ZKB-9 and 1.10.2.2-RXN had reference free energies of -40.52 and +40.52, respectively.

\end{outline}

\begin{table}[]
\begin{tabular}{|cllllll|}
\toprule
\multicolumn{1}{|c}{Compound} & \multicolumn{1}{c}{Formula} & \multicolumn{1}{c}{C} & \multicolumn{1}{c}{H} & \multicolumn{1}{c}{O} & \multicolumn{1}{c}{N} & \multicolumn{1}{c|}{\textbf{S}} \\
\toprule
ALA                 & C3H7NO2                              & 0.00                           & 1                              & -2                             & -3                             &                                \\
ARG                 & C6H14N4O2                            & 0.33                           & 1                              & -2                             & -3                             &                                \\
ASN                 & C4H8N2O3                             & 1.00                           & 1                              & -2                             & -3                             &                                \\
ASP                 & C4H7NO4                              & 1.00                           & 1                              & -2                             & -3                             &                                \\
CYS                 & C3H7NO2S                             & 0.67                           & 1                              & -2                             & -3                             & -2                             \\
GLN                 & C5H10N2O3                            & 0.40                           & 1                              & -2                             & -3                             &                                \\
GLU                 & C5H9NO4                              & 0.40                           & 1                              & -2                             & -3                             &                                \\
GLY                 & C2H5NO2                              & 1.00                           & 1                              & -2                             & -3                             &                                \\
HIS                 & C6H9N3O2                             & 0.67                           & 1                              & -2                             & -3                             &                                \\
ILE                 & C6H13NO2                             & -1.00                          & 1                              & -2                             & -3                             &                                \\
LEU                 & C6H13NO2                             & -1.00                          & 1                              & -2                             & -3                             &                                \\
LYS                 & C6H14N2O2                            & -0.67                          & 1                              & -2                             & -3                             &                                \\
MET                 & C5H11NO2S                            & -0.40                          & 1                              & -2                             & -3                             & -2                             \\
PHE                 & C9H11NO2                             & -0.44                          & 1                              & -2                             & -3                             &                                \\
PRO                 & C5H9NO2                              & -0.40                          & 1                              & -2                             & -3                             &                                \\
SER                 & C3H7NO3                              & 0.67                           & 1                              & -2                             & -3                             &                                \\
THR                 & C4H9NO3                              & 0.00                           & 1                              & -2                             & -3                             &                                \\
TRP                 & C11H12N2O2                           & -0.18                          & 1                              & -2                             & -3                             &                                \\
TYR                 & C9H11NO3                             & -0.22                          & 1                              & -2                             & -3                             &                                \\
VAL                 & C5H11NO2                             & -0.80                          & 1                              & -2                             & -3                             &                                \\
\hline
\textbf{Avg AA}  & C$_{5}$H$_{7.7}$O$_{1.5}$N$_{1.4}$S$_{0.04}$\ \cite{senko1995}  & 0.05  & 1  & -2  & -3  & -0.2 \\
&&&&&& \\
ADENINE             & C5H5N5                               & 2.00                           & 1                              &                                & -3                             &                                \\
GUANINE             & C5H5N5O                              & 2.40                           & 1                              & -2                             & -3                             &                                \\
THYMINE             & C5H6N2O2                             & 0.80                           & 1                              & -2                             & -3                             &                                \\
CYTOSINE            & C4H4N3O                              & 1.75                           & 1                              & -2                             & -3                             &                                \\
\hline
\textbf{Avg Nucleotide}            & C$_{19}$H$_{20}$O$_{4}$N$_{15}$                          & 1.74                           & 1                              & -2                             & -3                             &                               \\
\hline
\textbf{DNA}            & C$_{39}$H$_{44}$O$_{16}$N$_{15}$   & 0.85  & 1  & -2 & -3 & \\
\hline
\textbf{RNA} & C$_{38}$H$_{42}$O$_{20}$N$_{15}$ & 1.13 & 1  & -2  & -3 &\\
\hline
\textbf{Lipid} & C$_4$H$_{7.4}$O$_{1.35}$ & -1.18  & 1 & -2 &   &\\
\hline
\textbf{PHB} & C$_4$H$_{6}$O$_{2}$ & -0.50 & 1 & -2 & &\\
\bottomrule
\end{tabular}
\caption{Redox states of amino acids and nucleotides. Charge states of atoms were calculated with the python module OxidationNumberCalculator (https://github.com/Hiwen-STEM/OxidationNumberCalculator).}
\label{tab:atomic_charge_states}
\end{table}